\documentclass{svjour3}

\usepackage{pgfplotstable}
\usepackage{tabularx}
\usepackage{hyphenat}
\usepackage{xcolor}
\usepackage{paralist}
\usepackage[square,sort,comma,numbers]{natbib}
\usepackage{booktabs}
\usepackage{adjustbox}
\usepackage[linesnumbered, ruled, vlined]{algorithm2e}
\usepackage{pgfplots}
\usepackage{pifont}\newcommand{\cmark}{\ding{51}}
\newcommand{\xmark}{\ding{55}}
\usepackage{amsmath,amsfonts}
\usepackage{graphicx}
\usepackage{textcomp}
\usepackage{subfig}
\usepackage{float}
\usepackage{xcolor}
\usepackage[symbol]{footmisc}
\usepackage{makecell}
\usepackage{tcolorbox}
\usepackage{hyperref} 
\usepackage{color, colortbl}
\usepackage{multirow}\usepackage{mathrsfs}\usepackage[title]{appendix}\usepackage{textcomp}\usepackage{manyfoot}\usepackage{algorithmicx}\usepackage{algpseudocode}\usepackage{listings}\usepackage{rotating}
\usepackage{lscape}
\usepackage{siunitx}
\DeclareSIUnit{\million}{M}

\def\BibTeX{{\rm B\kern-.05em{\sc i\kern-.025em b}\kern-.08em
    T\kern-.1667em\lower.7ex\hbox{E}\kern-.125emX}}

\newcommand{\rev}[1]{{\color{black} #1}}
\newcommand{\minrev}[1]{{\color{black} #1}}

\begin{document}

\title{Impact of Log Parsing on Deep Learning-Based Anomaly Detection\footnote{This version of the article has been accepted for publication, after peer review  but is not the Version of Record and does not reflect post-acceptance improvements, or any corrections. The Version of Record is available online at: \url{https://doi.org/10.1007/s10664-024-10533-w}.}}

\author{Zanis Ali Khan   \and Donghwan Shin \and Domenico Bianculli  \and Lionel C. Briand 
}

\institute{Zanis Ali Khan  \at
              University of Luxembourg, Luxembourg, Luxembourg \\
              \email{zanis-ali.khan@uni.lu}          
           \and
             Donghwan Shin \at
             	University of Sheffield, Sheffield, United Kingdom  \\
             \email{d.shin@sheffield.ac.uk} 
            \and
             Domenico Bianculli \at
             University of Luxembourg, Luxembourg, Luxembourg \\
             \email{domenico.bianculli@uni.lu} 
	  \and
             Lionel C. Briand \at
             University of Ottawa, Canada, and the Lero SFI Centre for Software Research and University of Limerick, Ireland (Part of this work was done while the author was with the University of Luxembourg)\\
             \email{lbriand@uottawa.ca, lionel.briand@lero.ie} 
       }

\date{Received: date / Accepted: date}

\maketitle

\begin{abstract}
Software systems log massive amounts of data, recording
important runtime information. Such logs are used, for example, for log-based
anomaly detection, which aims to automatically detect abnormal
behaviors of the system under analysis by processing the information
recorded in its logs. Many log-based anomaly detection
techniques based on deep learning models include a pre-processing step called log parsing. However, understanding
the impact of log parsing on the accuracy of anomaly detection
techniques has received surprisingly little attention so far. Investigating what are the key properties log parsing techniques should ideally have to help anomaly detection is therefore warranted.

In this paper, we report on a comprehensive empirical study on the
impact of log parsing on anomaly detection accuracy, using 13 log
parsing techniques, \rev{seven} anomaly detection
techniques \rev{(five based on deep learning and two based on traditional machine learning)} on three publicly available log datasets. Our empirical
results show that, despite what is widely assumed, there is no strong correlation between log parsing accuracy and
anomaly detection accuracy, regardless of the metric used for
measuring log parsing accuracy. Moreover, we experimentally confirm
existing theoretical results showing that it is a property that we refer to as distinguishability in
log parsing results---as opposed to their accuracy---that plays an essential role in achieving accurate anomaly detection.

\keywords{Logs \and Log parsing \and Template identification \and Anomaly detection}
\end{abstract}

\section{Introduction}\label{sec:introduction}

Software system execution logs provide valuable information about the runtime 
behavior of the system, which is essential for monitoring and troubleshooting. 
Among many log analysis approaches, \textit{log-based anomaly detection} has 
been actively studied to automatically detect abnormal behaviors of the 
system under analysis by processing the information recorded in 
logs~\cite{10.1145/3460345}. Recently, anomaly detection techniques based on 
Deep Learning (DL) models, such as Long Short-Term Memory 
(LSTM)~\cite{du2017deeplog,meng2019loganomaly,zhang2019robust} and Convolutional 
Neural Networks (CNNs)~\cite{lu2018detecting}, have shown promising results.

One common aspect of most anomaly detection techniques is having a 
pre-processing step called \emph{log parsing} (also known as log template 
identification). This step is needed because anomaly detection techniques require structured 
logs to automatically process them, whereas input logs are often free-formed or 
semi-structured, as generated by logging statements (e.g., \texttt{printf()} 
and \texttt{logger.info()}) in the source code. Many log parsing 
techniques have also been developed to automatically convert unstructured input 
logs into structured logs~\cite{zhu2019tools}.

The frequent combination of log parsing and anomaly detection clearly implies the 
importance of the former for the latter. Nevertheless,
assessing in a systematic way the 
impact of log parsing on anomaly detection has received surprisingly little
attention so far. Only 
recently, \citet{shin2021theoretical} investigated what \textit{ideal} log parsing results 
are in terms of accurate anomaly detection, but purely from a theoretical standpoint. 
\citet{le2022log} empirically showed that different log parsing techniques, 
among other potential factors, can significantly affect anomaly detection 
accuracy, but the accuracy of log parsing results was not adequately measured, and 
the correlation between log parsing accuracy and anomaly detection accuracy 
was not reported. \rev{\citet{fu2023empirical} attempted to address the issue by evaluating log parsing and anomaly detection accuracy. However, they relied on a single log parsing accuracy metric~\cite{khan2022guidelines}, and the log parsing results used to evaluate anomaly detection techniques were based on less than 1\% of all logs used, which limits the validity of the findings.}

To systematically investigate the impact of log parsing on anomaly detection 
while addressing the issues of the aforementioned studies, this paper
reports on an empirical study, in which we performed a 
comprehensive evaluation using 13 log parsing techniques, \rev{seven  anomaly detection 
techniques---five based on deep learning and two based on traditional machine learning}---on \rev{three} publicly available log datasets. We considered all three log 
parsing accuracy metrics (i.e., grouping accuracy~\cite{zhu2019tools}, parsing 
accuracy~\cite{dai2020logram}, and template accuracy~\cite{khan2022guidelines}) 
proposed in the literature.

Against all assumptions, our results show that there is no strong correlation between log
parsing accuracy and anomaly detection accuracy, regardless of
the metric  used for measuring log parsing accuracy. In other words, 
accurate log parsing results do not necessarily increase anomaly detection 
accuracy. To better understand the phenomenon at play, 
we investigated another property of log parsing, \emph{distinguishability}, 
a concept proposed by \citet{shin2021theoretical} that was theoretically shown 
to relate to anomaly detection accuracy. Our empirical results confirm that,
as far as anomaly detection is concerned, 
distinguishability in log parsing results is the property 
that really matters and should be the key target of log parsing. 

In summary, the main contributions of this paper are:
\begin{itemize}
\item the systematic and comprehensive evaluation of the impact of log parsing 
on anomaly detection;
\item the investigation of the impact of the distinguishability of log parsing 
results on anomaly detection.
\end{itemize}

The rest of the paper is organized as follows. Section~\ref{sec:background} 
provides basic information used throughout the paper, including the definitions 
of logs, messages, and templates, as well as an overview of log parsing and 
anomaly detection. Section~\ref{sec:motivation} motivates our study and 
introduces the research questions. Section~\ref{sec:expr-design} describes 
the experimental design, including the log datasets, log parsing techniques, and 
anomaly detection techniques used in the experiments. Section~\ref{sec:results} 
presents the experimental results. Section~\ref{sec:implications} discusses the 
practical implications, derived from the results, for the application 
of log parsing in the context of anomaly detection. 
\rev{Section~\ref{sec:related-work} surveys the related work.} Section~\ref{sec:conclusion-future-work} concludes the paper and
provides directions for future 
work.

 \section{Background}\label{sec:background}

In this section, we provide an overview of the main concepts that will
be used throughout the paper. 
We first introduce the definitions of logs, messages, and log templates (\S~\ref{sec:pre}). 
We then explain the concept of log parsing (also known as log template
identification) and illustrate different log parsing accuracy metrics proposed in the literature (\S~\ref{sec:log-parsing}). 
We discuss log-based anomaly detection and the corresponding accuracy
metrics in \S~\ref{sec:anomaly-detection}. 
Finally, we summarize the recent theoretical results on ideal log
parsing for accurate anomaly detection, introducing the concept of \emph{distinguishability} for log parsing results (\S~\ref{sec:distinguishability}). 

\subsection{Logs, Messages, and Templates}\label{sec:pre}

A \emph{log} is a sequence of log entries\footnote{Note that a log is different 
from a log file. In practice, \emph{one} log file may contain \emph{many} logs 
representing the execution flows of different components/sessions. For example, 
an HDFS (Hadoop Distributed File System) log file contains many logs, 
distinguished by file block IDs, each representing an independent execution for 
a specific block.}. 
A \emph{log entry} contains various information about the event being
logged, including a timestamp, a logging level (e.g., \texttt{INFO}, \texttt{DEBUG}), and a log message. 
A \emph{log message} can be further decomposed into fixed and variable parts
since it is generated by executing a logging statement that can have
both fixed (hard-coded) strings and program variables in the source code.
For example, the execution of the logging statement 
``\texttt{logger.info("Deleting block " + blkID + " file " + fileName)}'' when the program variables 
\texttt{blkID} and \texttt{fileName} evaluate to \texttt{blk-1781} and \texttt{/hadoop/dfs}, respectively,
will generate a log entry ``\texttt{11:22:33 INFO Deleting block blk-1718 file /hadoop/dfs}'' 
where the log message ``\texttt{Deleting block blk-1718 file /hadoop/dfs}'' can be decomposed into
the fixed parts (i.e., ``\texttt{Deleting block}'' and ``\texttt{file}'') and the variable parts (i.e., ``\texttt{blk-1718}'' 
and ``\texttt{/hadoop/dfs}'').
A \emph{(log message) template} masks the various elements of each
variable part with a special character ``\texttt{<*>}''; this
representation is widely used in log-based analyses (e.g., log 
parsing~\cite{he2017drain,jiang2008abstracting}, anomaly 
detection~\cite{zhang2019robust,du2017deeplog}, and log-based 
testing~\cite{elyasov2012log, jeong2020log}) when it is important to focus on 
the event types captured by a log message.

For instance, the template corresponding to the example log message 
``\texttt{Deleting block blk-1178 file /hadoop/dfs}'' is ``\texttt{Deleting block
  <*> file <*>}''.

\subsection{Log Parsing (Log Template Identification)}\label{sec:log-parsing}
Although software execution logs contain valuable information about 
the run-time behavior of the software system under analysis, 
they cannot be directly processed by log-based analysis techniques that require
structured input logs (containing templates) instead of free-formed log messages.
Extracting log templates from log messages is straightforward 
when the source code with the corresponding logging statements is available. 
However, often the source code is unavailable, for example, due to the
usage of 3rd-party, 
proprietary components. This leads to the problem of log parsing (log template identification):
\textit{How can we identify the log templates of log messages without accessing the source code?}

To address this problem, many automated log-parsing approaches, 
which take as input log messages and identify their log templates using different 
heuristics, have been proposed in the literature (e.g., AEL~\cite{jiang2008abstracting}, 
Drain~\cite{he2017drain}, IPLoM~\cite{makanju2009clustering}, 
LenMa~\cite{shima2016length}, LFA~\cite{nagappan2010abstracting},
LogCluster~\cite{7367331}, LogMine~\cite{hamooni2016logmine}, Logram~\cite{dai2020logram}, 
LogSig~\cite{tang2011logsig}, MoLFI~\cite{messaoudi2018search},
SHISO~\cite{mizutani2013incremental}, SLCT~\cite{vaarandi2003data},
and Spell~\cite{du2016spell}). 

Three different accuracy metrics have been proposed to evaluate the accuracy of log parsing 
approaches: \textit{Grouping Accuracy} (GA)~\cite{zhu2019tools}, \textit{Parsing 
Accuracy} (PA)~\cite{dai2020logram}, and \textit{Template Accuracy} 
(TA)~\cite{khan2022guidelines}. 

\citet{zhu2019tools} observed that log parsing can be considered as a clustering 
process where log messages with the same template are clustered into the same 
group. Based on this idea, they proposed the GA metric to assess if log messages 
are correctly grouped. Specifically, GA is defined as the ratio of log messages 
\textit{correctly parsed} by the log parsing approach under evaluation over the 
total number of log messages, where a log message is correctly parsed when its 
log message group is the same as the ground truth (i.e., a group generated by 
oracle templates). 

\citet{dai2020logram} later proposed PA, to address the issue that GA only considers 
message groups, not the equivalence between the templates identified by the log 
parsing approach under evaluation and the oracle templates. Although having 
correctly grouped messages would be enough in some cases (e.g., detecting 
anomalies based on the sequence of template IDs without considering the content 
of the templates~\cite{du2017deeplog}), correctly identified templates (i.e., 
templates identical to the corresponding oracle ones) matter when the fixed parts of 
templates are used (e.g., detecting anomalies based on the semantic information 
in the templates~\cite{zhang2019robust}). To this end, PA replaces the 
definition of a correctly parsed log message in GA as follows: a log message is 
\textit{correctly parsed} when its identified template is identical to the 
oracle template. 

\citet{khan2022guidelines} recently proposed the TA metric, since both GA and PA are 
defined based on the number of correctly parsed log messages and, therefore, can 
be misleading, especially when there are many repeated messages (e.g., heartbeat 
messages). Specifically, they introduced Precision-TA (PTA) and Recall-TA (RTA), 
where PTA is defined as the number of correctly identified templates over the 
total number of identified templates and RTA is defined as the number of 
correctly identified templates over the total number of oracle templates. 
Moreover, FTA (short for ``F1-measure TA'') is the harmonic mean of PTA and RTA.

\subsection{Anomaly Detection}\label{sec:anomaly-detection}
(Log-based) anomaly detection is a technique that aims to identify
anomalous patterns, recorded in input logs, that do not conform to the
expected behaviors of the system under
analysis~\cite{10.1145/3460345}. It takes as input a sequence of log
templates and determines whether the given sequence represents a
normal behavior of the system or not.

With the recent advances in Deep Learning (DL), many anomaly detection
approaches, which leverage DL models to learn various aspects of log template sequences of normal and abnormal behaviors and classify them, have been  proposed in the literature; for example, 
DeepLog~\cite{du2017deeplog}, LogAnomaly~\cite{meng2019loganomaly}, and 
LogRobust~\cite{zhang2019robust} are based on Long Short-Term Memory based (LSTM), 
CNN~\cite{lu2018detecting} is based on Convolutional Neural Network,
and PLELog~\cite{yang2021semi} is based on Gated recurrent units (GRUs). 

To assess the accuracy of anomaly detection approaches,
it is common practice to use standard metrics 
from the information retrieval domain, such as \emph{Precision}, \emph{Recall}, and \emph{F1-Score}. 
These metrics are defined as follows: 
$\textit{Precision} = \frac{\textit{TP}}{\textit{TP} + \textit{FP}}$,
$\textit{Recall} = \frac{\textit{TP}}{\textit{TP} + \textit{FN}}$, and
$\textit{F1-score} = \frac{2\times \textit{Precision} \times \textit{Recall}}{\textit{Precision} + \textit{Recall}}$
where \textit{TP} (True Positive) is the number of abnormal logs correctly identified by the model,
\textit{FP} (False Positive) is the number of normal logs incorrectly identified as anomalies by the model, and
\textit{FN} (False Negative) is the number of abnormal logs incorrectly identified as normal.

\subsection{Ideal Log Parsing Results for Accurate Anomaly Detection}\label{sec:distinguishability}

Given the dependency of anomaly detection on log parsing, 
\citet{shin2021theoretical} presented a theoretical analysis on 
\emph{ideal} log parsing results for accurate anomaly detection. 
The idea behind the analysis is that log parsing can be regarded as the abstraction 
of log messages, where some tokens in the messages are converted to variable parts. 
Then, if normal and abnormal logs are over-abstracted by log parsing so that 
they are \textit{indistinguishable} from each other, 
it is clear that anomaly detection, which takes as input the parsed logs 
(i.e., abstracted logs, sequences of templates), cannot distinguish normal from abnormal logs. 
Based on this idea, they formally defined the concept of \emph{distinguishability} 
\rev{as a property of} log parsing results and showed that it is an essential condition for ideal log parsing results.

Specifically, let $M$ be a set of log messages and $L$ be a set of
logs where a log $l\in L$ is a sequence of log messages $\langle \mathit{m_1,
m_2}, \dots, m_n \rangle$.
Also, let $L_n \subseteq L$ be a set of normal logs and $L_a \subseteq L$ 
be a set of abnormal logs such that $L_n \cap L_a = \emptyset$ and $L_n \cup L_a = L$. 
Given $M$ and a set of templates (i.e., log parsing results) $T$, 
an abstraction function $\tau: M \to T$ that represents a generic log parsing approach can be defined.
Based on $\tau$, an abstraction of a log $l = \langle \mathit{m_1, m_2}, \dots, m_n \rangle$ 
can be defined as $\tau^*(l) = \langle \mathit{ \tau(m_1), \tau(m_2)}, \dots, \tau(m_n) \rangle$. 
Similarly, an abstraction of a set of logs $L$ can be defined as $\tau^{**}(L) = \{\tau^*(l) \mid l \in L\}$.
Notice that $\tau^{**}(L)$ represents a log parsing result for a set of logs $L$.

The notion of distinguishability can be defined as follows: 
$\tau$ \emph{distinguishes} $L_n$ and $L_a$ if and only if $\tau^{**}(L_n) \cap \tau^{**}(L_a) = \emptyset$.
In other words, a log parsing approach distinguishes between normal and abnormal logs 
if and only if they are still distinguishable after log parsing.
When $\tau$ distinguishes $L_n$ and $L_a$, $\tau^{**}(L)$ for $L = L_n \cup L_a$ is 
called \emph{d-maintaining}, meaning that the distinguishability between 
$L_n$ and $L_a$ is \emph{maintained} in the log parsing result.  \section{Motivation}\label{sec:motivation}

As discussed in Section~\ref{sec:background}, log parsing converts unstructured 
logs into structured ones, which  can then be processed by log-based
analysis techniques like anomaly detection.
It is quite natural to speculate that log 
parsing results can affect anomaly detection results.
Intuitively, the research literature has assumed that inaccurate log 
parsing results leads to inaccurate anomaly detection results.
However, this hypothesis has not been fully investigated in the
literature, except for one empirical study~\cite{le2022log} and one 
analytical investigation~\cite{shin2021theoretical}. 

\citet{le2022log} recently presented an empirical work 
investigating several aspects that can impact Deep Learning (DL)-based 
anomaly detection approaches, such as 
training data selection, data grouping, class distribution, data noise, and early detection ability.
One of their experiments considering data noise assessed the impact of noise deriving 
from log parsing results. Specifically, they used four log parsing techniques (Drain~\cite{he2017drain}, 
Spell~\cite{du2016spell}, AEL~\cite{jiang2008abstracting}, and IPLoM~\cite{makanju2009clustering}) 
to generate log parsing results for two log datasets (BGL~\cite{oliner2007supercomputers} and
Spirit~\cite{oliner2007supercomputers}). 
Then, for each log dataset, they used the different log parsing
results as input of  five anomaly detection approaches (DeepLog~\cite{du2017deeplog}, LogAnomaly~\cite{meng2019loganomaly}, 
PLELog~\cite{yang2021semi}, LogRobust~\cite{zhang2019robust}, and 
CNN~\cite{lu2018detecting}), and measured the accuracy of the latter.
Their experimental results showed that log parsing approaches highly influence 
the accuracy of anomaly detection; 
for example, the F1-Score of DeepLog on Spirit logs~\cite{oliner2007supercomputers}
decreases from $0.755$ to $0.609$ when Drain is used instead of IPLoM for log parsing. 

Although this is the first clear evidence showing the impact of log parsing 
results on anomaly detection accuracy, the scope of the underlying
study is limited. For example, it simply uses different log parsing
results (produced by different tools)
without quantitatively assessing the accuracy of the log parsing tools; 
therefore, the relationship between log parsing accuracy and anomaly detection accuracy remains unclear. 
To this end, we define our first research question as follows: 
\textbf{\emph{RQ1 - To which extent does the accuracy of log parsing affect the accuracy of anomaly detection?}}

As summarized in Section~\ref{sec:distinguishability}, \citet{shin2021theoretical} recently proposed a theoretical 
framework determining the ideal log parsing results for anomaly
detection by introducing the concept of ``distinguishability'' for
log parsing results. It is argued that, rather than accuracy as previously assumed, what really matters is the extent to which log parsing results are distinguishable. 
However, to the best of our knowledge, there is no empirical work
assessing quantitatively distinguishability in log parsing results
and its impact on anomaly detection accuracy. 
Therefore, we define our second research question as follows:
\textbf{\emph{RQ2 - How does the 
accuracy of anomaly detection vary with distinguishability of log parsing 
results?}}

Answering the above questions will have a significant impact on both research 
and industry in the field of log-based anomaly detection. For example, if the 
answer to the first question is that, regardless of the log parsing accuracy 
metrics, there is no relationship between log parsing accuracy and anomaly 
detection accuracy, then it means that there is no need to use the existing 
accuracy metrics to evaluate log parsing results for anomaly detection. This 
would completely change the way log parsing tools are evaluated. Similarly, if 
the answer to the second question is that the distinguishability of log parsing 
results indeed affects anomaly detection, as expected from the recent 
theoretical analysis~\cite{shin2021theoretical}, then this must be the focus of log parsing evaluations. As a result, our answers will 
provide essential insights on better assessing the quality of log parsing techniques
for more accurate anomaly detection.

 \section{Experimental Design}\label{sec:expr-design}

All experiments presented in this paper were carried out using 
the HPC facilities of the University of Luxembourg (see https://hpc.uni.lu).
Specifically, we used Dual Intel Xeon Skylake CPU (8 cores) and 64GB RAM
for running individual log parsing and anomaly detection techniques. 

\subsection{Datasets}\label{sec:datasets}

To answer the research questions introduced in Section~\ref{sec:motivation}, we used publicly available
datasets based on the LogHub benchmark~\cite{he2020loghub}, which contains 
a large collection of log messages from various types of systems including operating systems 
(Linux, Windows, and Mac), distributed systems (BGL, Hadoop, HDFS, Thunderbird, and 
OpenStack),  
standalone programs (Proxifier and Zookeeper), and mobile systems (Android). 
The benchmark has been widely used in various studies focused on 
log parsing~\cite{khan2022guidelines, zhu2019tools, dai2020logram} and
anomaly detection~\cite{le2022log, fu2023empirical}.

\begin{table}
\centering
\caption{Datasets in LogHub benchmark~\cite{he2020loghub}}
\label{table:datasets_literature}
\small

\begin{tabular}{lrr}

\toprule
      Datasets & Anomaly Label & Source Code  \\
 \midrule
 
 Android & \xmark &  \xmark \\
  Apache & \xmark &  \xmark  \\
   BGL & \cmark & \xmark \\
   \rowcolor{lightgray}
 HDFS &  \cmark &  \cmark  \\
  HPC & \xmark &  \xmark  \\
   \rowcolor{lightgray}
 Hadoop &  \cmark & \cmark \\
 HealthApp & \xmark &  \xmark  \\
 Linux & \xmark &  \xmark \\
 Mac & \xmark & \xmark \\
  OpenSSH & \xmark & \xmark  \\
   \rowcolor{lightgray}
 OpenStack &  \cmark &  \cmark   \\
 Proxifier & \xmark &  \xmark  \\
 Spark & \xmark & \xmark  \\
 Spirit & \cmark & \xmark  \\
 Thunderbird & \cmark & \xmark    \\
 Windows & \xmark &  \xmark \\
 Zookeeper & \xmark &  \xmark \\
 
\bottomrule
\end{tabular}

 \end{table}

Among the benchmark datasets, we selected HDFS, Hadoop, and OpenStack datasets
because of the following reasons: 
\begin{inparaenum}[(1)]
\item they have labels for normal and abnormal logs to be used for assessing the accuracy of anomaly detection techniques \textit{and} 
\item the source code of the exact program version used to generate
  the logs is  publicly available; this allows us to extract correct oracle templates 
(i.e., ground truth templates) for each log message. 
\end{inparaenum}
The oracle templates are especially important in our study as we need to 
carefully assess both log parsing accuracy and anomaly detection accuracy.
Although the benchmark provides some oracle templates for all log datasets, they are \textit{manually 
generated} (without accessing the source code) and cover \textit{only}  2K log messages randomly sampled for 
each dataset. As discussed by \citet{khan2022guidelines}, those manually generated oracle 
templates are \textit{error-prone}; therefore, we used the logging statements in the source code to extract 
correct oracle templates. 
Table~\ref{table:datasets_literature} shows all the log datasets in the LogHub benchmark
and whether they meet each of the above-mentioned criteria; the rows
highlighted in gray meet both criteria. 

During our preliminary evaluation, we found an issue with HDFS.
The original HDFS logs were too large \rev{(\SI{11.2}{\million} log messages)} to be processed by the
slowest anomaly detection technique (i.e.,
LogAnomaly~\cite{meng2019loganomaly}) when setting a two-day timeout. 
Due to the large number of experiments we needed to conduct (i.e., all
combinations of log parsing and anomaly detection techniques with
additional repeats for distinguishable and indistinguishable log
parsing results, see \S~\ref{sec:eval-rq1-method} and \S~\ref{sec:eval-rq2-method}), 
we decided to reduce the log dataset size.
As we found that the
slowest log parsing technique (i.e., LogAnomaly) could process up to $n=300K$ messages
within 2 hours, we randomly and iteratively removed logs (i.e., sequences of log messages)
from the HDFS dataset to reduce it until the total number of remaining messages 
was less than 300K. 
Notice that each HDFS log is a sequence of log messages having the same block ID,
representing either a normal or abnormal sequence of events.
To preserve individual (normal or abnormal) sequences, 
we randomly selected and removed them by sequence, not by message. 
\rev{Although the resulting reduced dataset is much smaller than the original dataset, it is still representative of the original dataset in terms of the distribution of normal and abnormal log messages. Specifically, the original HDFS dataset consists of \num{11175629} log messages, with 97.43\% normal and 2.57\% abnormal log messages, and the reduced HDFS dataset mirrors this distribution, with 97.60\% normal and 2.40\% abnormal log messages.}

\begin{table}
\centering
\caption{Size information of the log datasets used in our
  experiments. Number of oracle templates (\textit{O});
Number of all logs ($L_\mathit{all}$); Number of normal logs ($L_\mathit{n}$);  Number of abnormal logs ($L_\mathit{a}$); Number of all messages ($M_\mathit{all}$); Number of messages in normal logs ($M_\mathit{n}$); Number of messages in abnormal logs ($M_\mathit{a}$).}
\label{table:datasets}
\small
\begin{tabular}{l@{\hspace{2pt}}r@{\hspace{5pt}}r@{\hspace{5pt}}r@{\hspace{5pt}}r@{\hspace{5pt}}r@{\hspace{5pt}}r@{\hspace{5pt}}r@{\hspace{5pt}}}

\toprule
      Dataset & O & $L_{all}$ & $L_{n}$ & $L_{a}$ & $M_{all}$ & $M_{n}$ & $M_{a}$  \\
 \midrule
HDFS (reduced)  & 26 &  15295 & 15026 & 269 & 299971 & 292776 & 7195    \\
Hadoop &  175 & 54 & 11 & 43 & 109968 & 14392 & 95576  \\
OpenStack & 21 &  2068 & 2064  & 4 & 79925 & 79817 & 108   \\
\bottomrule
\end{tabular}

 \end{table}

Table~\ref{table:datasets} reports on the size of our
datasets, in terms of the number of oracle templates (\textit{O}), the
number of all logs ($L_\mathit{all}$), the number of normal logs ($L_\mathit{n}$), the number of abnormal logs ($L_\mathit{a}$), the number of all messages ($M_\mathit{all}$), the number of messages in normal logs ($M_\mathit{n}$), and the number of messages in abnormal logs ($M_\mathit{a}$). Note that the number of log messages is the same as the number of log entries (see Section~\ref{sec:pre} for details).

 \subsection{Log Parsing Techniques}\label{sec:log-parsing-techniques}

We aimed to use as many log parsing techniques as possible, among those available in the literature.
Since \citet{khan2022guidelines} recently provided a comprehensive evaluation 
of 14 log parsing techniques (i.e., 
AEL~\cite{jiang2008abstracting}, Drain~\cite{he2017drain}, 
IPLoM~\cite{makanju2009clustering}, LenMa~\cite{shima2016length}, 
LFA~\cite{nagappan2010abstracting}, LKE~\cite{fu2009execution}, 
LogCluster~\cite{7367331}, LogMine~\cite{hamooni2016logmine}, Logram~\cite{dai2020logram}, LogSig~\cite{tang2011logsig}, 
MoLFI~\cite{messaoudi2018search}, SHISO~\cite{mizutani2013incremental}, 
SLCT~\cite{vaarandi2003data}, and Spell~\cite{du2016spell}),
we decided to reuse their replication package, including all the
aforementioned techniques.

However, we had to exclude LKE since our preliminary evaluation results
showed that it could not complete its run for \textit{all} of our log datasets
within the 2-day timeout. 
Notice that we have already reduced our log datasets (in particular, HDFS), as discussed in Section~\ref{sec:datasets},
based on the slowest anomaly detection technique (i.e., LogAnomaly). 
Although we could additionally reduce the datasets based on the slowest log parsing technique (i.e., LKE),
we found that it would result in small logs that are not representative of the size and complexity of real-world logs.

As a result, we considered 13 log parsing techniques in our experiments.
For all the log parsing techniques, we used their default parameters.
 \subsection{Anomaly Detection Techniques}\label{sec:anomaly-detection-techniques}

Similar to the case of log parsing techniques, we considered the work
of \citet{le2022log}, 
a recent empirical study that evaluated five DL-based anomaly detection techniques 
(i.e.,  DeepLog~\cite{du2017deeplog}, LogAnomaly~\cite{meng2019loganomaly}, 
LogRobust~\cite{zhang2019robust}, PLELog~\cite{yang2021semi}, and
CNN~\cite{lu2018detecting}), and decided to use their replication
package, including all the aforementioned techniques. For all anomaly detection techniques, we 
used their default parameters. 
These techniques are representative of the state of the art of
DL-based anomaly detection techniques.

\rev{In addition to deep learning models, we included two representative traditional machine learning models,
namely Support Vector Machine (SVM)~\cite{hearst1998support} and Random Forest (RF)~\cite{breiman2001random}\footnote{We used the implementations from the scikit-learn library~\cite{pedregosa2011scikit}.}
since they are known for their effectiveness in anomaly detection tasks on the HDFS
dataset~\cite{wu2023effectiveness, jia2023robust}.
}

\rev{We want to note that the seven anomaly detection techniques used in this paper \emph{all require log parsing as a preliminary step}. 
Although a few recent studies~\cite{le2021log, mvula2023heart, nedelkoski2020selfAtt}  have proposed anomaly detection techniques that do not require log parsing, we did not consider them in our work.
This is mainly because our focus is on assessing the impact of log parsing on anomaly detection techniques.
We leave the evaluation of techniques that do not require log parsing for future work.}

\subsection{Methodology for RQ1}\label{sec:eval-rq1-method}
\rev{Recall that RQ1 investigates to what extent the accuracy of log parsing affects the accuracy of anomaly detection.}
To answer RQ1, for each dataset, we first executed the log parsing techniques to
generate log parsing results and computed their accuracy in terms of GA, 
PA, and FTA (see 
\S~\ref{sec:log-parsing}). We then executed the anomaly
detection techniques on each of the log parsing results and computed their
accuracy in terms of precision (PR), recall (RE), and F1 score.
By doing so, we obtained a tuple of accuracy values $\langle \mathit{GA,
PA, FTA, PR, RE, F1} \rangle$ for each combination of datasets, log parsing
results, and anomaly detection techniques.

For log parsing, we executed each of the log parsing techniques with a
2-day timeout. Since MoLFI is non-deterministic, we executed it three
times. In total, we obtained 16 log parsing results (three from the
three different executions of MoLFI and 13 from the
remaining log parsing techniques) for each dataset. For each log parsing result,
we computed $\langle \mathit{GA, PA, FTA} \rangle$ using the oracle templates (and the
messages matching them) for the corresponding datasets.

For anomaly detection, we divided the individual log parsing results into two
disjoint sets, i.e., a training set and a test set, using a split ratio of 80:20.
\rev{Considering the data leakage problem mentioned by \citet{le2022log}, 
we used the first 80\% of the logs (in chronological order) for training and the remaining 20\% for testing.}
We trained the anomaly detection techniques on each of the training sets with a
2-day timeout, and used the corresponding test sets to compute
$\langle \mathit{PR, RE, F1} \rangle$. To account for the randomness of anomaly detection
techniques, we repeated the train-and-test process five times and used the average 
F1 score.

As a result, we obtained \rev{224} tuples $\langle \mathit{GA,
PA, FTA, PR, RE, F1} \rangle$ from the
combinations of two datasets, 16 log parsing results, and \rev{seven} anomaly detection
techniques.

\subsection{Methodology for RQ2}\label{sec:eval-rq2-method}
\rev{Recall that RQ2  investigates the relationship between the distinguishability of log parsing results and anomaly detection accuracy.}
To answer RQ2, we need distinguishable and indistinguishable log parsing results to compare in terms 
of anomaly detection accuracy. Although the log parsing results generated for RQ1 are available, they 
are mostly \rev{(but not all)} distinguishable, leading to unbalanced data for RQ2. 
To systematically assess the impact of the distinguishability of log
parsing results on anomaly detection accuracy using balanced
data, we generate pairs of distinguishable and indistinguishable log
parsing results.

Specifically, let $d(R)$ be the distinguishability --- expressed as a Boolean value,
either \emph{true} ($T$) or \emph{false} ($F$) --- of a log 
parsing result $R$. For each log parsing result $R$ 
\rev{(i.e., the result of executing a log parsing technique for a dataset)}
generated in the context of 
RQ1 (i.e., 16 log parsing results for each of the two datasets), we first 
created a pair of log parsing results $\langle R, R' \rangle$ by
artificially generating $R'$ 
from $R$ such that $d(R') = \neg d(R)$ using Algorithms~\ref{alg:indist} and \ref{alg:dist}, detailed further below. 
By definition, if $R$ is distinguishable then 
$R'$ will be indistinguishable and vice versa.  
For the sake of simplicity, we denote the distinguishable result (be
it $R$ or $R'$) as $R_\mathit{dst}$ and the indistinguishable one
(respectively, either $R'$ or $R$) as $R_\mathit{ind}$. 
We then executed, for all pairs $\langle R_\mathit{dst}, R_\mathit{ind} \rangle$, 
all the considered anomaly detection techniques twice: the 
first time using $R_\mathit{dst}$ as input and the second time using $R_\mathit{ind}$ as 
input; for each run of each anomaly detection technique we computed its accuracy in terms of 
precision, recall, and F1 score. By doing so, we obtained the anomaly detection 
accuracy scores for pairs of distinguishable ($R_\mathit{dst}$) and 
indistinguishable ($R_\mathit{ind}$) versions of log parsing results,
and then compared them. 

For the generation of $R'$ from $R$, it is important to minimize the difference 
between $R$ and $R'$ \rev{(in terms of both training and testing datasets)} while achieving $d(R') = \neg d(R)$. 
This is to ensure that if there is a difference in anomaly detection scores between $R$ and $R'$, it is mostly due to distinguishability and not to other differences between $R$ and $R'$ (e.g., the number of templates or the size of log parsing results). 
\rev{Furthermore, the testing datasets for $R$ and $R'$ should remain the same.}
To do this, we need to distinguish the two cases when $d(R) = T$ and when $d(R) = F$, as described below.

\subsubsection{Generation of Indistinguishable from Distinguishable Log Parsing Results}\label{sec:ind-from-dst}

When $d(R) = T$ (i.e., $R = R_\mathit{dst}$), it means that 
templates for different log messages in $R$ are different enough to distinguish 
between normal and abnormal logs in $R$, as explained in 
Section~\ref{sec:distinguishability}. 
For example, let us consider two logs $l_1 = \langle \mathit{m_1, m_2} \rangle$ and $l_2 = \langle \mathit{m_3, m_4} \rangle$ where the templates of the four messages are identified as $\tau(m_1) = t_1$, $\tau(m_2) = t_2$, $\tau(m_3) = t_3$, and $\tau(m_4) = t_2$, respectively, using a log parsing technique $\tau$. 
Figure~\ref{fig:dist-example} shows the logs, messages, and templates. 
In this case, the log parsing result of $\tau$ for $\{l_1, l_2\}$ is 
\textit{distinguishable}, as highlighted in blue in the figure, since
$\tau^*(l_1) = \langle \tau(m_1), \tau(m_2) \rangle = \langle t_1,
t_2 \rangle$ and $\tau^*(l_2) = \langle \mathit{\tau(m_3), \tau(m_4)}
\rangle = \langle t_3, t_2 \rangle$ are different 
(due to $\tau(m_1) \neq \tau(m_3)$, i.e., $t_1 \neq t_3$). However, if the templates of $m_1$ and 
$m_3$ were the same, then the log parsing result would be 
\textit{indistinguishable}. In other words, as highlighted in red in the figure, we can make the distinguishable log 
parsing result of $\tau$ indistinguishable by merging the templates of $m_1$ and $m_3$ 
(e.g., by introducing a dummy log parsing technique $\tau'$ that
behaves the same as $\tau$ 
except for $\tau'(m_1) = \tau'(m_3) = t_{13}$).
Notice that $\tau'$ changes only (a few) templates, not the corresponding log messages,
meaning that the original datasets remain the same.
Using this idea, to generate $R' = R_\mathit{ind}$ from $R = R_\mathit{dst}$, we 
generated the templates of $R_\mathit{ind}$ by iteratively merging the templates of $R_\mathit{dst}$ 
until $d(R_\mathit{ind}) = F$. 
Furthermore, to minimize the difference between $R_\mathit{dst}$ and $R_\mathit{ind}$
in terms of the number of templates (i.e., to minimize the number of templates 
being merged), we start with merging the templates with the highest number of 
matching messages in the log. This is based on the intuition that the more 
messages affected by merging templates, the more likely normal and abnormal 
logs are to become indistinguishable. 
\rev{Recall that we only change the templates, not their log messages.}

\rev{Although merging templates to generate indistinguishable log parsing results might look artificial,
it is indeed realistic to some extent.
In practice, a log parsing result would be indistinguishable only when a log parsing technique fails to identify proper templates that can sufficiently ``distinguish'' normal and abnormal log sequences. 
Therefore, merging templates in the distinguishable log parsing
results mimics the behavior of such imperfect log parsing techniques, leading to indistinguishable log parsing results.}

One might \rev{also} object that artificially merging templates  corresponding to 
different messages could introduce incorrect templates in $R_\mathit{ind}$, 
leading to an unfair comparison between $R_\mathit{dst}$ and $R_\mathit{ind}$. 
However, it is common for the log parsing techniques to identify many templates 
that are already incorrect~\cite{khan2022guidelines}. Furthermore,
the focus of RQ2 is not 
 the correctness of templates but rather the distinguishability of log 
parsing results. Our goal is to generate a pair of $R_\mathit{dst}$ and 
$R_\mathit{ind}$ that are as similar as possible except for the
distinguishability property.  
\rev{Indeed, the testing datasets for $R_\mathit{dst}$ and 
$R_\mathit{ind}$ are the same in terms of log messages and their order. The only difference 
lies in how individual log messages are mapped to the templates, affecting the distinguishability of log parsing results. 
Consequently, the only difference between $R_\mathit{dst}$ and 
$R_\mathit{ind}$ is in their distinguishability, ensuring that no bias is introduced when evaluating the model's performance.}

Algorithm~\ref{alg:indist} summarizes the above-mentioned idea into
the pseudocode for generating $R_\mathit{ind}$ from
$R_\mathit{dst}$. After initializing $R_\mathit{ind}$
(line~\ref{alg:indist:init}) as a copy of $R_\mathit{dst}$, the algorithm extracts the set of templates $T$ of $R_\mathit{dst}$ (line~\ref{alg:indist:getT}) and sorts the templates in $T$ in ascending order by the number of matching messages (line~\ref{alg:indist:sort}). The algorithm then iteratively merges the last $n$ templates (starting from $n=2$ as initialized at line~\ref{alg:indist:int}) in the sorted templates list $T_s$ (i.e., merging the top-$n$ templates that have the highest number of matching templates) until $R_\mathit{ind}$ becomes indistinguishable (lines~\ref{alg:indist:while-start}--\ref{alg:indist:while-end}). Notice that the while loop does not continue endlessly since $R_\mathit{ind}$ must be indistinguishable when $n$ becomes $|T_s|$ (i.e., all templates are merged into one) by definition. The algorithm ends by returning $R_\mathit{ind}$. 

\begin{algorithm}
\SetNoFillComment
\SetKwInOut{Input}{Input}
\SetKwInOut{Output}{Output}

\Input{Distinguishable Log Parsing Result $R_\mathit{dst}$}
\Output{Indistinguishable Log Parsing Result $R_\mathit{ind}$} 
\BlankLine

Log Parsing Result (Set of Parsed Logs) $R_\mathit{ind} \gets \textit{copy}(R_\mathit{dst})$\label{alg:indist:init}\\
Set of Templates $T \gets \textit{getTemplates}(R_\mathit{dst})$\label{alg:indist:getT}\\
Sorted List of Templates $T_s \gets \textit{sortByNumMessages}(T)$\label{alg:indist:sort}\\
Integer $n \gets 2$\label{alg:indist:int}\\
\While{$d(R_\mathit{ind}) = \textit{True}$}{\label{alg:indist:while-start}
    Set of Templates $T_m \gets \textit{getLastTemplates}(T_s, n)$\label{alg:indist:getT-to-merge}\\
    $R_\mathit{ind} \gets \textit{mergeTemplates}(T_m, R_\mathit{dst})$\label{alg:indist:merge}\\
    $n \gets n + 1$\label{alg:indist:inc}\\
}\label{alg:indist:while-end}
\textbf{return} $R_\mathit{ind}$\label{alg:indist:return}\\

\caption{Generating an indistinguishable log parsing result from a
  distinguishable one}
\label{alg:indist}
\end{algorithm}

\begin{figure}
\small
\begin{adjustbox}{width=\textwidth}
\begin{tabularx}{\textwidth}{lccc}
\toprule

Log & Template  &  Parsed Log (original)  & Parsed Log (after mer- \\ 
 & & & -ging $t_1$ and $t_3$ into $t_{13}$) \\
\midrule

$l1$ = $\langle m_1, m_2 \rangle$ & $\tau(m_1) = t_1$ , $\tau(m_2) = t_2$ & \color{teal}$\tau^*(l_1) = \langle t_1, t_2 \rangle$ & \color{orange}$\tau'^*(l_1) = \langle t_{13}, t_2 \rangle$  \\

\midrule

$l2$ = $\langle m_3, m_4 \rangle$ & $\tau(m_3) = t_3$ , $\tau(m_4) = t_2$ & \color{teal}$\tau^*(l_2) = {\langle t_3, t_2 \rangle}$ & \color{orange}$\tau'^*(l_2) = \langle t_{13}, t_2 \rangle$ \\

\bottomrule 
\end{tabularx}

 \end{adjustbox}
\caption{An example of making a \textcolor{teal}{distinguishable} log parsing result \textcolor{orange}{indistinguishable} by merging templates}
\label{fig:dist-example}
\end{figure}

\subsubsection{Generation of Distinguishable from Indistinguishable Log Parsing Results}\label{sec:dst-from-ind}

When $d(R) = F$ (i.e., $R = R_\mathit{ind}$), although one 
could do the dual of merging templates (i.e., dividing templates), it would require to
determine which templates to divide and how many templates to generate from a given template. Instead, we
adopted another 
heuristic: we removed the normal (or abnormal) logs that are indistinguishable 
from abnormal (or normal) logs. This is based on our observation that, 
when $d(R) = F$, only a small number of normal and abnormal logs are 
indistinguishable. To minimize the impact of removing logs, we removed normal 
logs when the total number of normal logs is larger than that of abnormal logs 
(as it is the case for the HDFS dataset); otherwise, we removed
abnormal logs (in the case of the Hadoop dataset). 
\rev{Specifically, only MoLFI, SLCT, LogCluster, and LFA generated indistinguishable log parsing results for HDFS in the first place, 
and we only removed 5, 5, 9, and 2 logs, respectively, out of 15026 normal logs.}

Algorithm~\ref{alg:dist} shows how to generate $R_\mathit{dst}$ from $R_\mathit{ind}$ based on the above idea. It first extracts the set of indistinguishable logs $L_\mathit{ind}$ from $R_\mathit{ind}$ (line~\ref{alg:dist:indist}). It then removes either normal or abnormal logs in $L_\mathit{ind}$ from $R_\mathit{ind}$ to generate $R_\mathit{dst}$ depending on the total number of normal and abnormal logs (lines~\ref{alg:dist:checkSize}--\ref{alg:dist:rma}). Since $R_\mathit{dst}$ is the result of removing indistinguishable (normal or abnormal) logs from $R_\mathit{ind}$, $R_\mathit{dst}$ is distinguishable. The algorithm ends by returning $R_\mathit{dst}$. 

\begin{algorithm}
\SetNoFillComment
\SetKwInOut{Input}{Input}
\SetKwInOut{Output}{Output}

\Input{Indistinguishable Log Parsing Result $R_\mathit{ind}$}
\Output{Distinguishable Log Parsing Result $R_\mathit{dst}$} 
\BlankLine

Set of Indistinguishable Logs $L_\mathit{ind} \gets \textit{getIndistLogs}(R_\mathit{ind})$\label{alg:dist:indist}\\
\If{$\textit{numNormalLogs}(R_\mathit{ind}) \geq \textit{numAbnormalLogs}(R_\mathit{ind})$}{\label{alg:dist:checkSize}
    Set of Parsed Logs $R_\mathit{dst} \gets R_\mathit{ind} \setminus \textit{getNormalLogs}(L_\mathit{ind})$\label{alg:dist:rmn}\\
}
\Else{
    Set of Parsed Logs $R_\mathit{dst} \gets R_\mathit{ind} \setminus \textit{getAbnormalLogs}(L_\mathit{ind})$\label{alg:dist:rma}\\
}
\textbf{return} $R_\mathit{dst}$\label{alg:dist:return}\\

\caption{Generating a distinguishable log parsing result from an
  indistinguishable one}
\label{alg:dist}
\end{algorithm}

\subsubsection{Treatment for Anomaly Detection Techniques using Semantic Information of Templates}\label{sec:treat-for-semantic}
Some of the anomaly detection techniques (i.e., LogRobust~\cite{zhang2019robust}, PLELog~\cite{yang2021semi}, LogAnomaly~\cite{meng2019loganomaly}) use the semantic information of 
templates, instead of simply using template IDs, by converting them into 
semantic vectors~\cite{jurafsky2019vector}. 
\rev{For these techniques, two templates are considered ``identical'' if their semantic vectors are similar enough.}
Therefore, the notion of ``identical'' templates for 
determining the distinguishability of log parsing results must be revised in 
terms of the semantic vectors used by these anomaly detection techniques\rev{; otherwise, 
simply determining the distinguishability based on their template IDs would be meaningless for these techniques.}
To do this, for each log parsing result $R$, we applied a clustering algorithm to the 
semantic vectors of all templates and considered the 
templates in the same cluster to be identical. Specifically, we used 
DBSCAN~\cite{backlund2011density} for clustering since it does not require the 
number of clusters as an input parameter. 
For instance, in the above example $\tau$ with $m_1$ and $m_3$, 
if the semantic vectors of $\tau(m_1)$ and $\tau(m_3)$ belong to the 
same cluster, then the templates of $m_1$ and $m_3$ are considered the same. 
\rev{Note that the semantic vectors are carefully designed to capture subtle semantic 
nuances and are able to identify semantically similar log templates while distinguishing 
different ones~\cite{zhang2019robust}. Therefore, clustering these semantic vectors can 
effectively identify ``identical'' templates for the semantic-based anomaly detection 
techniques.}
We then followed the same heuristics described above to generate 
$R'$ from $R$ based on the clustered templates.

\subsubsection{\rev{Additional Analysis: Degree of Distinguishability}}\label{sec:dist-degree}

\rev{
So far, we have described how to compare distinguishable and indistinguishable log parsing results to 
answer RQ2, treating distinguishability as a binary property (i.e., either distinguishable or 
indistinguishable) following the original definition~\cite{shin2021theoretical}. 
Although we have effectively minimized the difference between distinguishable and 
indistinguishable log parsing results to make a fair comparison, we have applied an \textit
{artificial} process for generating indistinguishable log parsing results from distinguishable 
ones (or vice versa).
To address this limitation, we present an additional analysis on the degree of 
distinguishability of the log parsing results generated for RQ1.

However, defining a metric to measure the degree of distinguishability is not straightforward, 
mainly because the original definition of distinguishability is too strict; for example, the 
log parsing result of two log sequences representing the same behavior can be considered 
distinguishable simply when they are different in length.
Therefore, we present a metric to measure the degree of distinguishability based on 
the number of common templates between normal and abnormal log sequences. 
This is based on the observation that a higher number of shared templates between normal and 
abnormal log sequences indicates weaker distinguishability.

Specifically, recall that we can consider a log parsing result $\tau^{**}(L)$ of a set of log 
sequences $L$ for a log parsing technique $\tau$.
Let $c(\tau^{**}(L))$ be the number of unique templates in $\tau^{**}(L)$.
We define the distinguishability score $\textit{distScore}(\tau, L)$ of $L$ for $\tau$ as the 
ratio of the number of common templates generated by $\tau$ between normal and abnormal log 
sequences to the number of unique templates in all log sequences in $L$, 
i.e., $\textit{distScore}(\tau, L) = 1 - \frac{c(\tau^{**}(L_n) \cap \tau^{**}(L_a))}{c(\tau^{**}(L))}$, 
where $L_n$ and $L_a$ are the sets of normal and abnormal log sequences in $L$, respectively.
Since $c(\tau^{**}(L)) = c(\tau^{**}(L_n)) \cup c(\tau^{**}(L_a))$, 
the distinguishability score is effectively the Jaccard distance between $L_n$ and $L_a$ in terms of their templates.
For example, the number of unique templates identified by Drain for the HDFS dataset is 31. 
Among them, 13 templates appear in both normal and abnormal log sequences. 
Therefore, the distinguishability score of Drain for the HDFS dataset is $1 - \frac{13}{31} = 0.57$.

We want to note that, ideally speaking, this additional analysis should allow us to measure 
the impact of distinguishability on anomaly detection accuracy in a more fine-grained manner 
without generating artificial log parsing results.
However, our metric is a heuristic and may not fully capture the various aspects 
of distinguishability.
Therefore, we will use this new analysis as a complementary study to the 
main analysis (treating distinguishability as a binary property), 
to provide a more comprehensive understanding of 
the impact of distinguishability on anomaly detection accuracy.
}

 \section{Results}\label{sec:results}

\subsection{RQ1: Relationship between Log Parsing Accuracy and Anomaly Detection Accuracy}\label{sec:eval-rq1-results}

All 13 log parsing techniques and \rev{7} anomaly detection techniques completed their executions on the HDFS and Hadoop datasets. 
However, none of the anomaly detection techniques detected abnormal logs in the OpenStack dataset (i.e., the F1 score is zero). 
This could be due to the very small number of abnormal logs in the dataset 
(only 4 out of 2068, as reported in Table~\ref{table:datasets}).
Therefore, we disregard the results for OpenStack. 

\begin{figure}
\centering

\begin{tabularx}{\textwidth}{
        l X  @{\hspace{3pt}}  X@{\hspace{1pt}}  l 
  }
    & \multicolumn{1}{c}{\textbf{GA}}
    & \multicolumn{1}{c}{\textbf{PA}}
    & \multicolumn{1}{c}{\textbf{FTA}}\\
    
    & \raisebox{-0.5\height}{\includegraphics[width=0.30\textwidth] {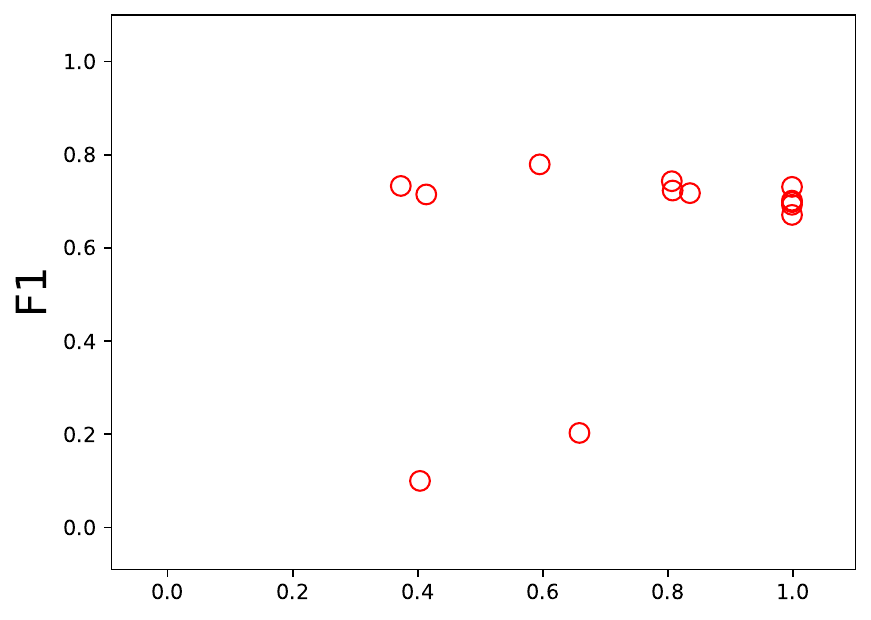}}
    & \raisebox{-0.5\height}{\includegraphics[width=0.29\textwidth] {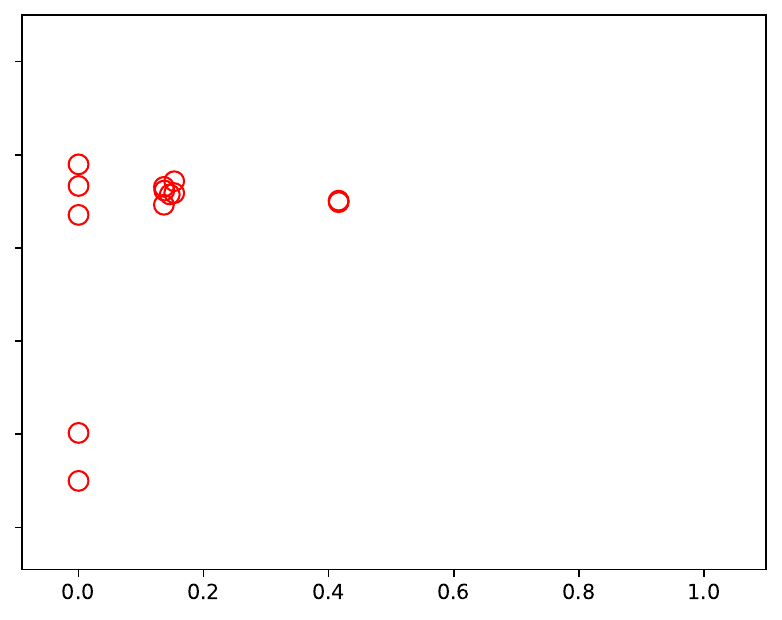}}
     & \raisebox{-0.5\height}{\includegraphics[width=0.29\textwidth] {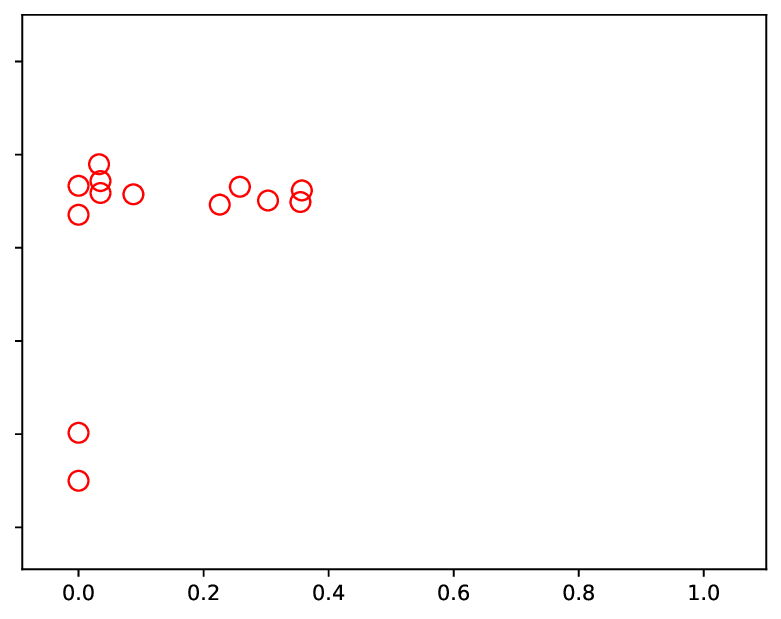}}
       \rotatebox[origin=c]{270}{\textbf{DeepLog}} \\
    
    &  \raisebox{-0.5\height}{\includegraphics[width=0.30\textwidth] {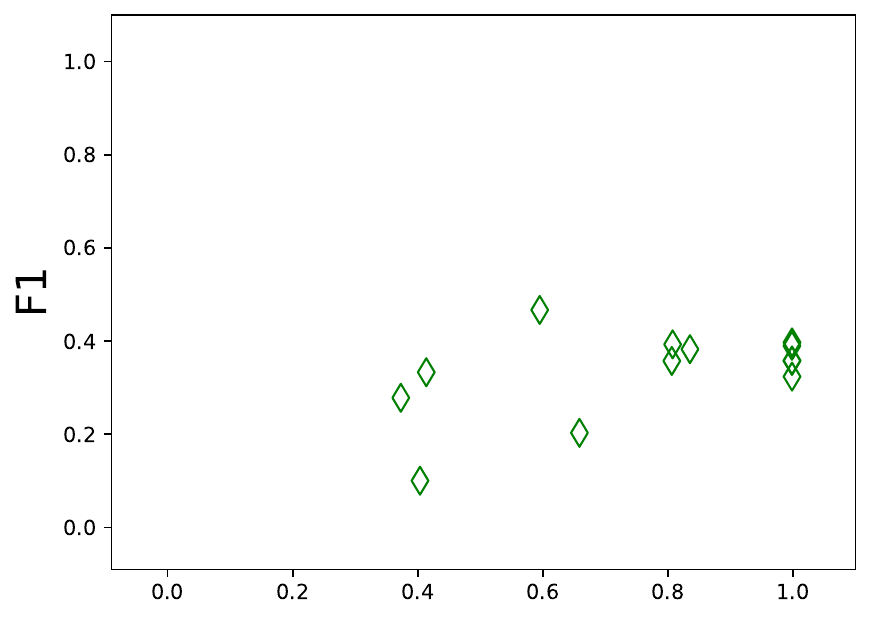}}
    &  \raisebox{-0.5\height}{\includegraphics[width=0.29\textwidth] {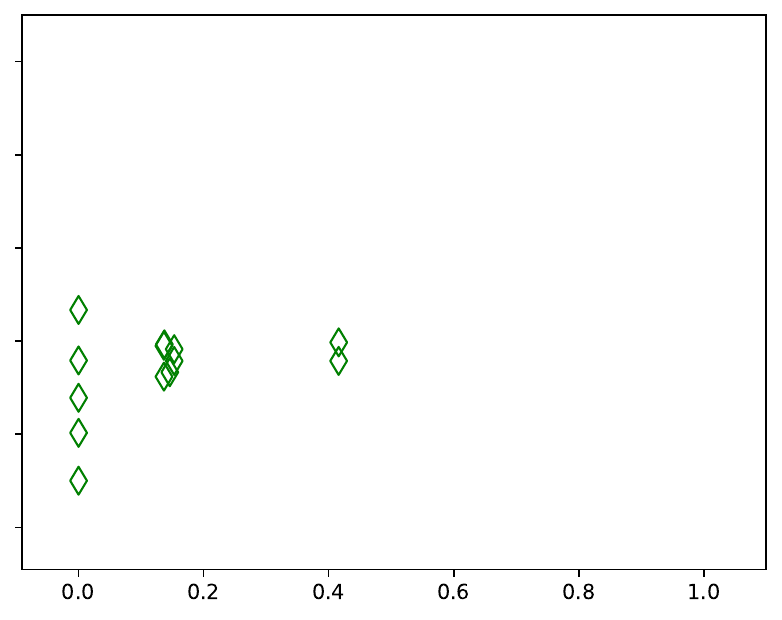}}
    &  \raisebox{-0.5\height}{\includegraphics[width=0.29\textwidth] {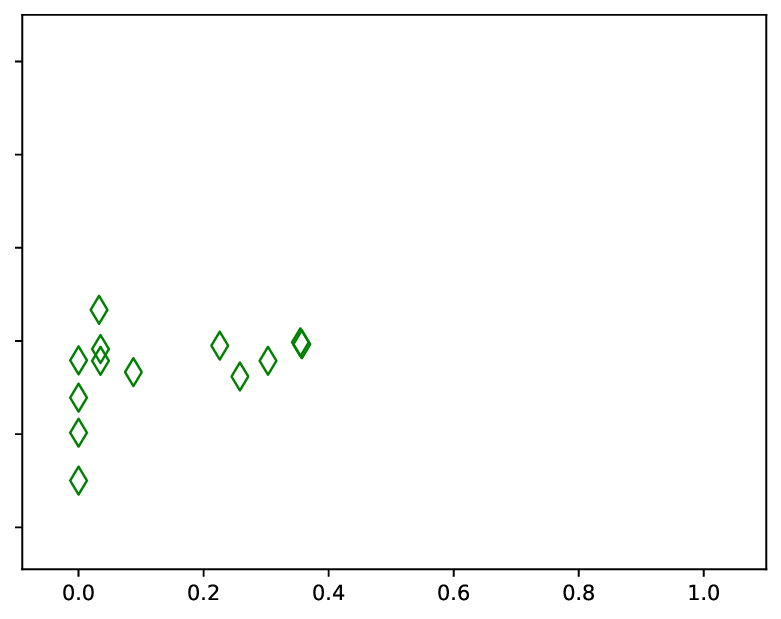}}
    \rotatebox[origin=c]{270}{\textbf{LogAnomaly}} \\

    & \raisebox{-0.5\height}{\includegraphics[width=0.30\textwidth] {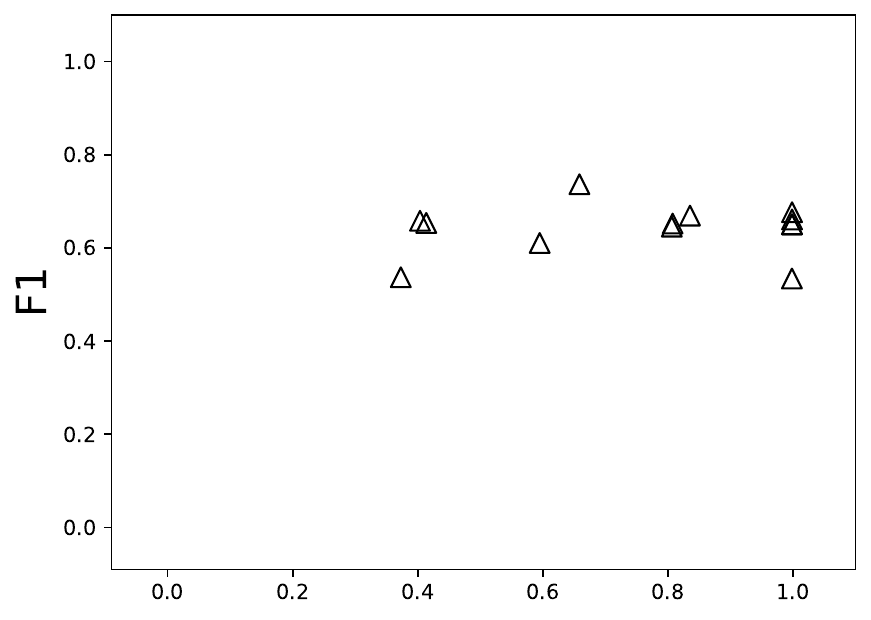}}
    & \raisebox{-0.5\height}{\includegraphics[width=0.29\textwidth] {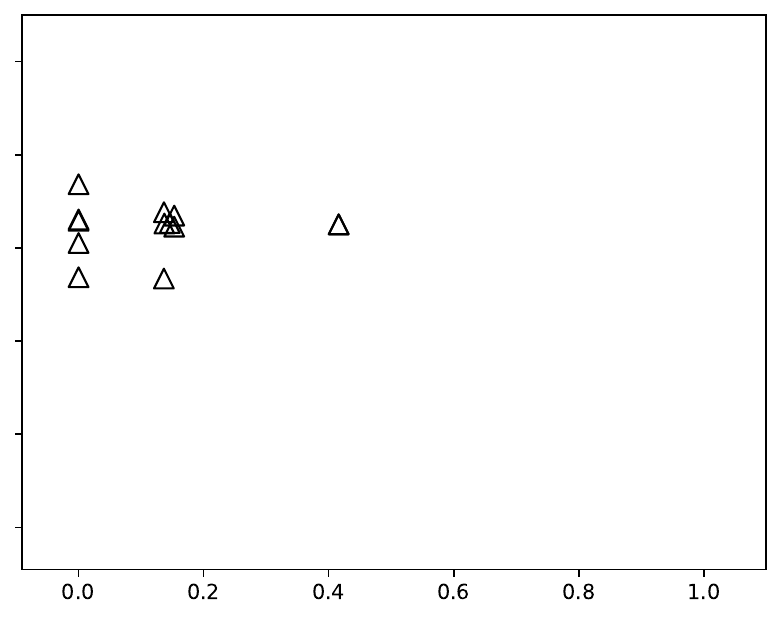}}
    & \raisebox{-0.5\height}{\includegraphics[width=0.29\textwidth] {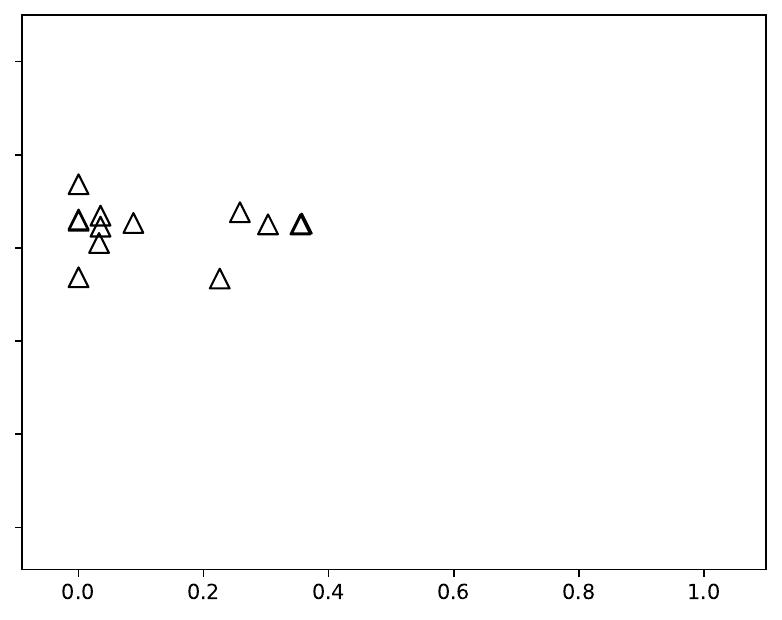}}
    \rotatebox[origin=c]{270}{\textbf{LogRobust}} \\

    & \raisebox{-0.5\height}{\includegraphics[width=0.30\textwidth] {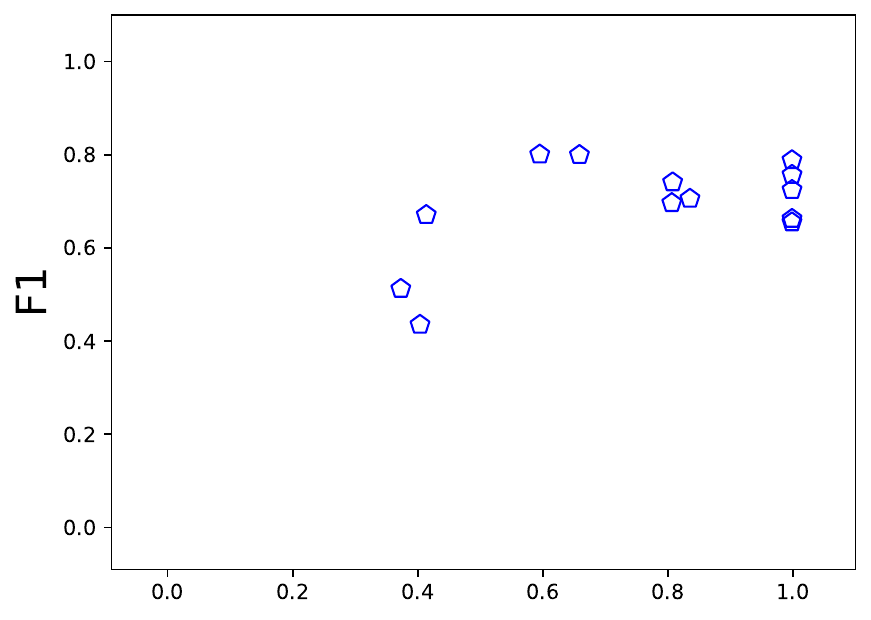}}
    & \raisebox{-0.5\height}{\includegraphics[width=0.29\textwidth] {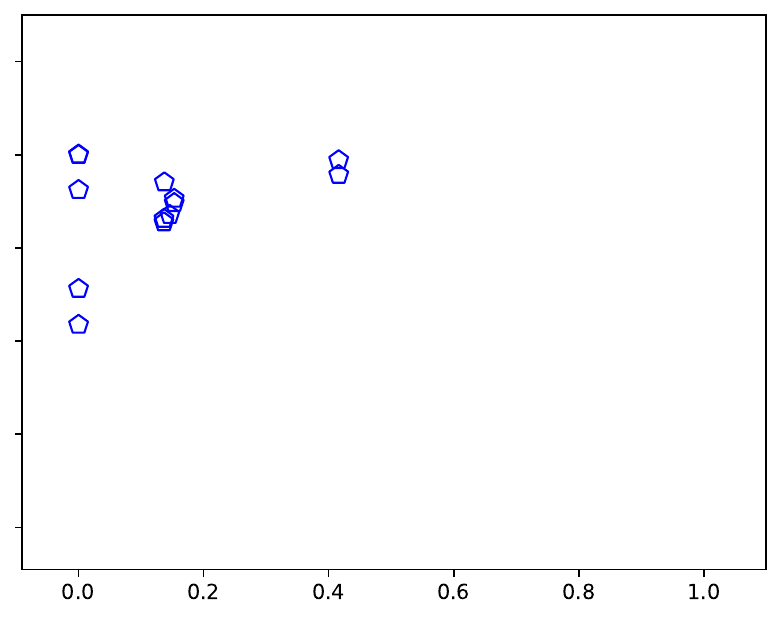}}
    & \raisebox{-0.5\height}{\includegraphics[width=0.29\textwidth] {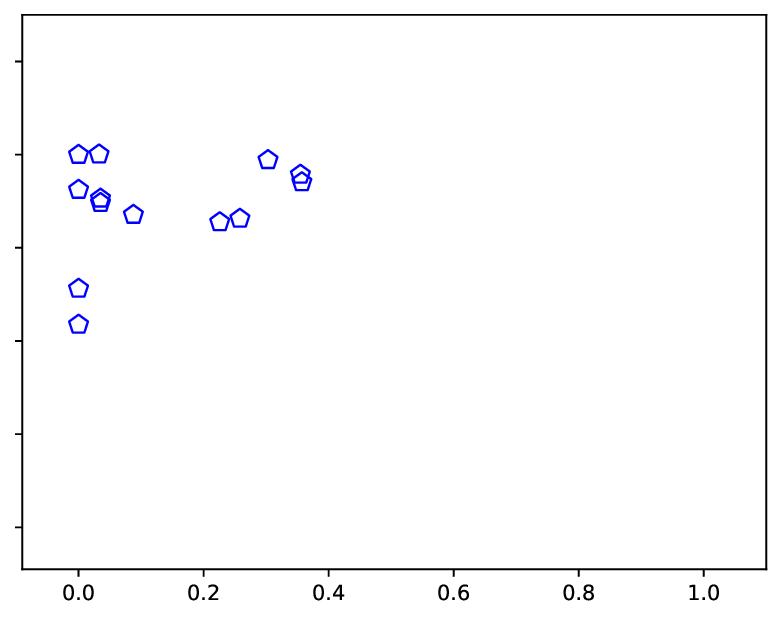}}
    \rotatebox[origin=c]{270}{\textbf{CNN}} \\

    & \raisebox{-0.5\height}{\includegraphics[width=0.30\textwidth] {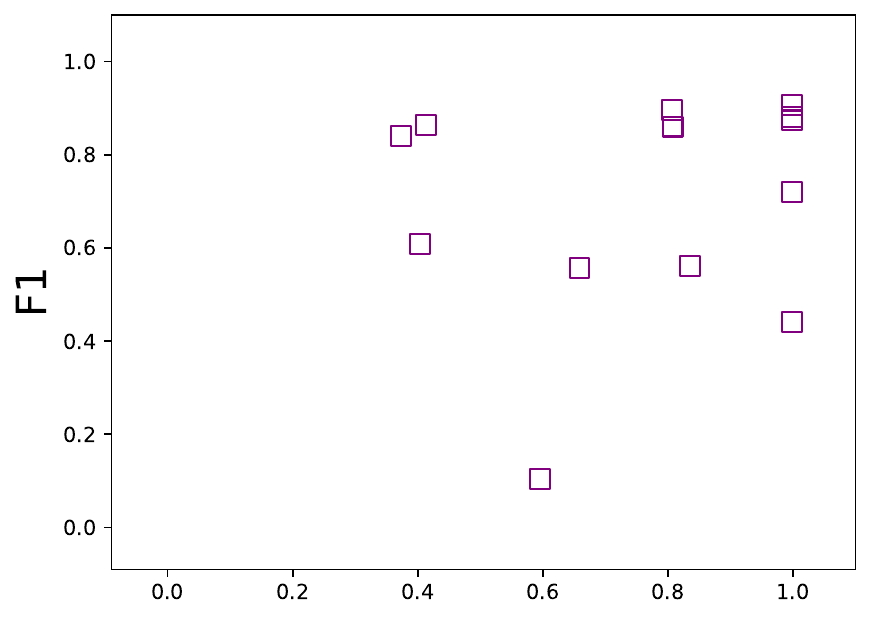}}
    & \raisebox{-0.5\height}{\includegraphics[width=0.29\textwidth] {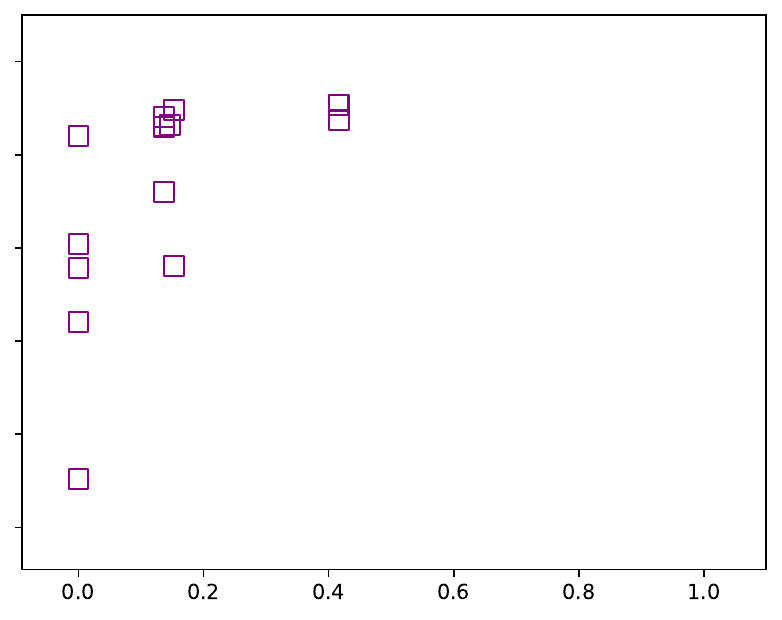}}
    & \raisebox{-0.5\height}{\includegraphics[width=0.29\textwidth] {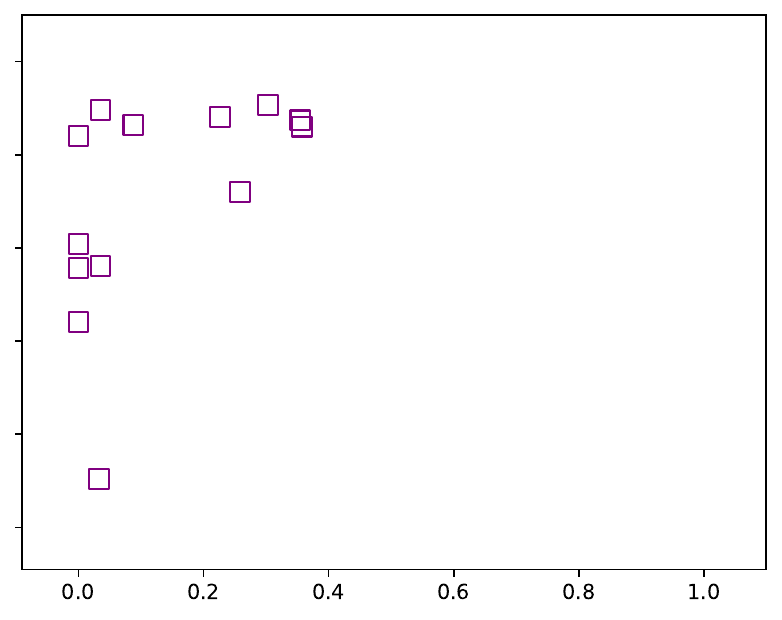}}
    \rotatebox[origin=c]{270}{\textbf{PLELog}} \\
    
     & \raisebox{-0.5\height}{\includegraphics[width=0.30\textwidth] {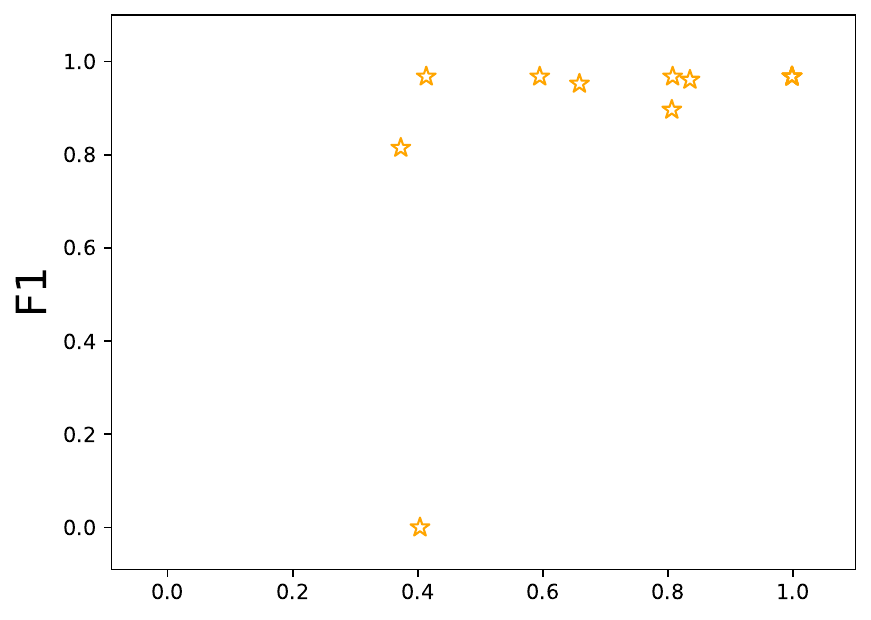}}
    & \raisebox{-0.5\height}{\includegraphics[width=0.29\textwidth] {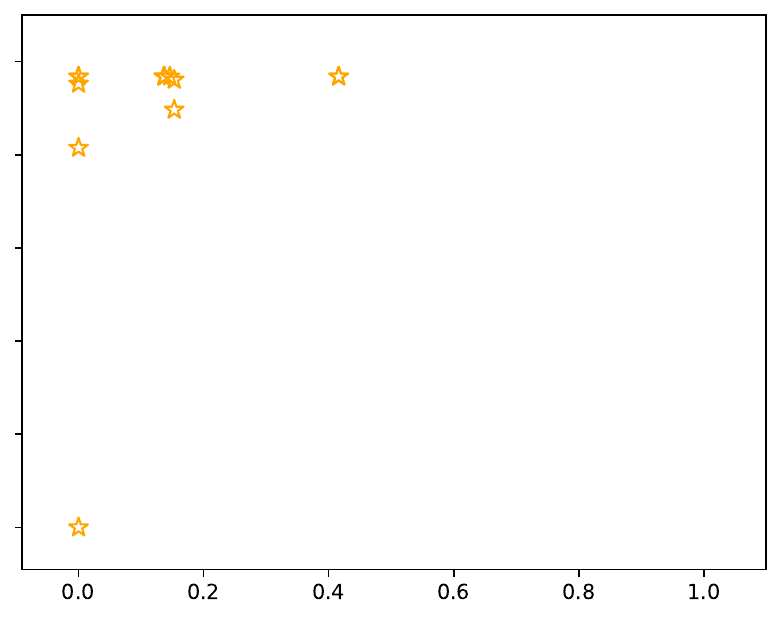}}
    & \raisebox{-0.5\height}{\includegraphics[width=0.29\textwidth] {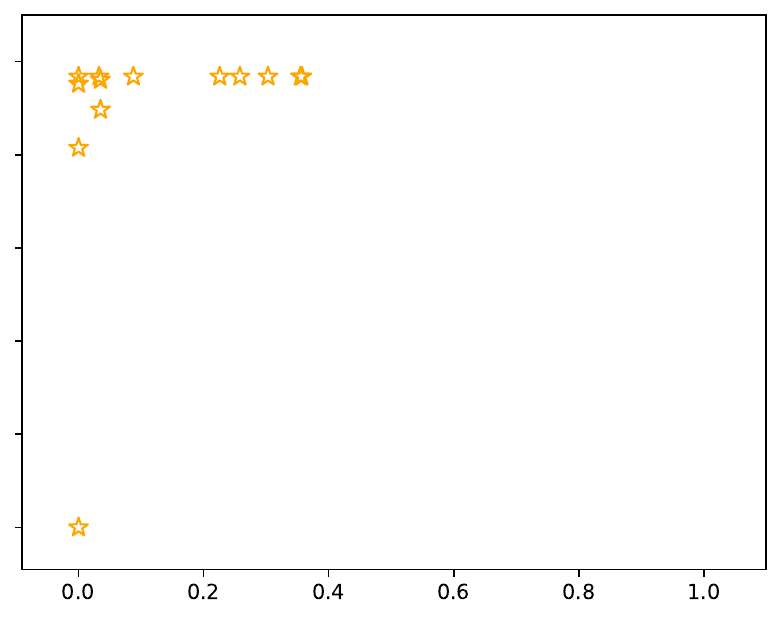}}
    \rotatebox[origin=c]{270}{\textbf{SVM}} \\

     & \raisebox{-0.5\height}{\includegraphics[width=0.30\textwidth] {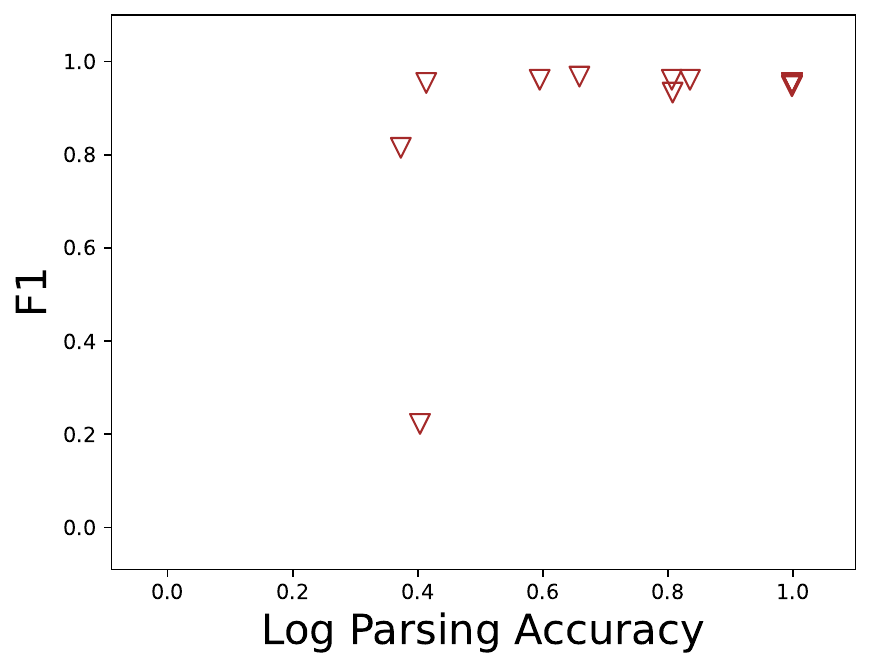}}
    & \raisebox{-0.5\height}{\includegraphics[width=0.29\textwidth] {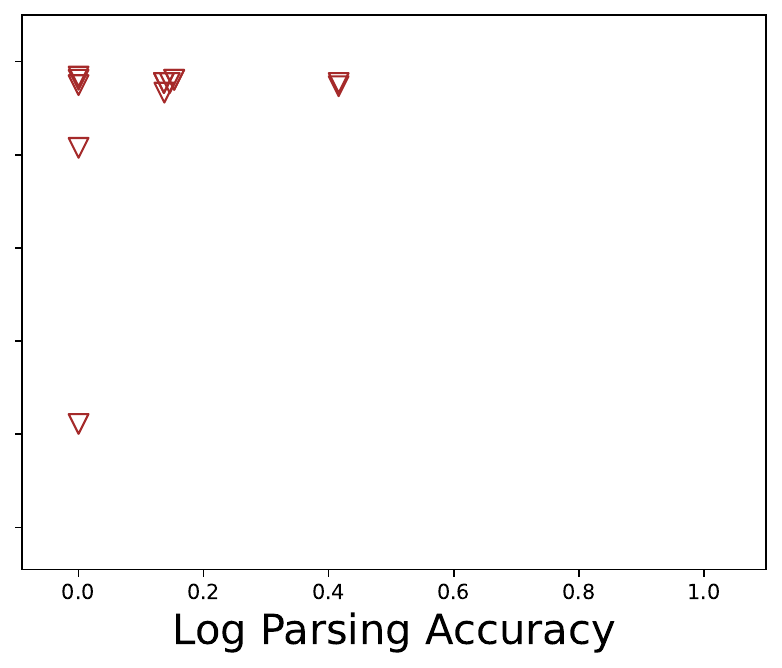}}
    & \raisebox{-0.5\height}{\includegraphics[width=0.29\textwidth] {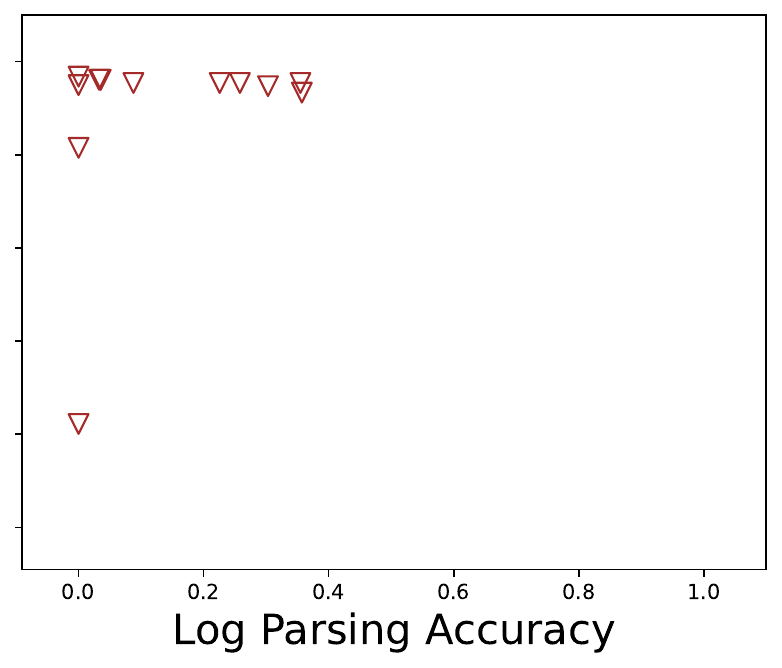}}
    \rotatebox[origin=c]{270}{\textbf{RF}} \\
    
\end{tabularx}

\caption{Relationship between TI accuracy and AD accuracy (HDFS)
}
\label{fig:rq1-hdfs-results}
\end{figure}

\begin{figure}
\centering

\begin{tabularx}{\textwidth}{
        l X  @{\hspace{3pt}}  X@{\hspace{1pt}}  l 
  }
    & \multicolumn{1}{c}{\textbf{GA}}
    & \multicolumn{1}{c}{\textbf{PA}}
    & \multicolumn{1}{c}{\textbf{FTA}}\\
    
    & \raisebox{-0.5\height}{\includegraphics[width=0.30\textwidth] {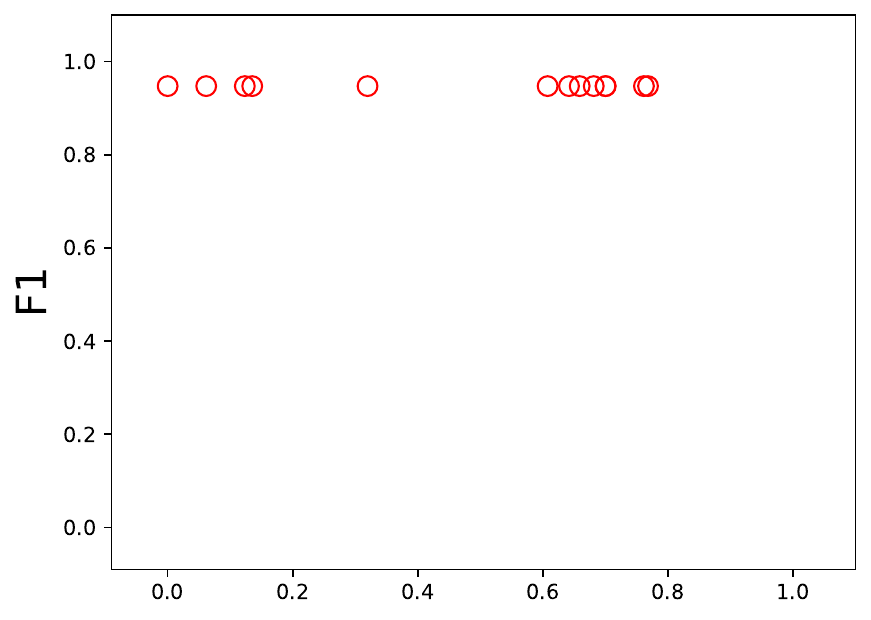}}
    & \raisebox{-0.5\height}{\includegraphics[width=0.29\textwidth] {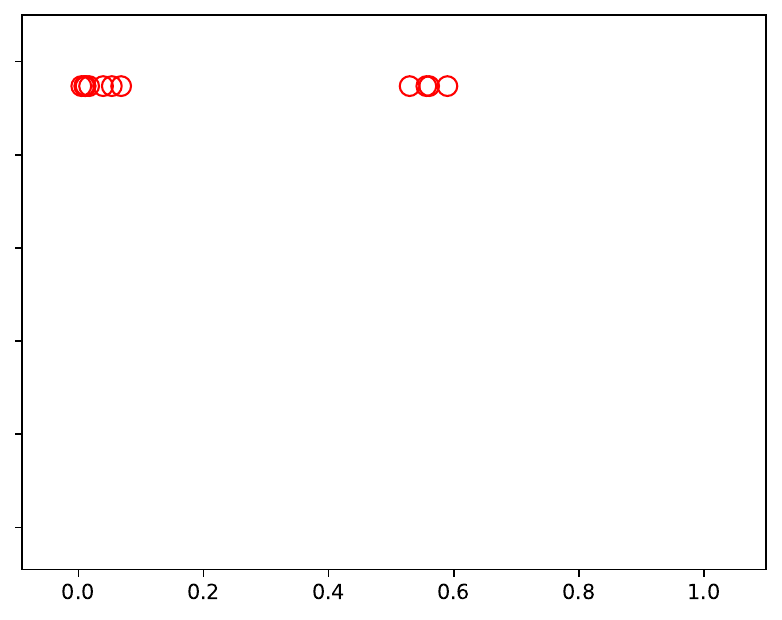}}
     & \raisebox{-0.5\height}{\includegraphics[width=0.29\textwidth] {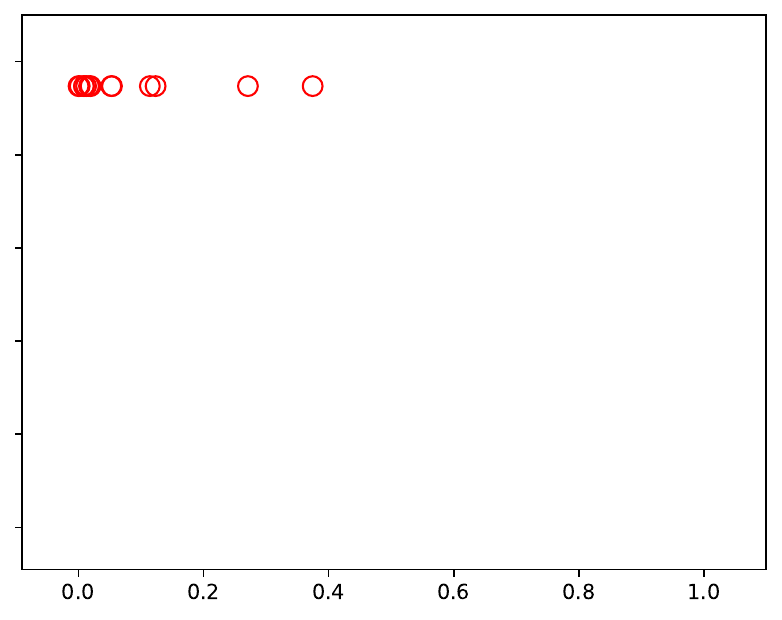}}
       \rotatebox[origin=c]{270}{\textbf{DeepLog}} \\
    
    &  \raisebox{-0.5\height}{\includegraphics[width=0.30\textwidth] {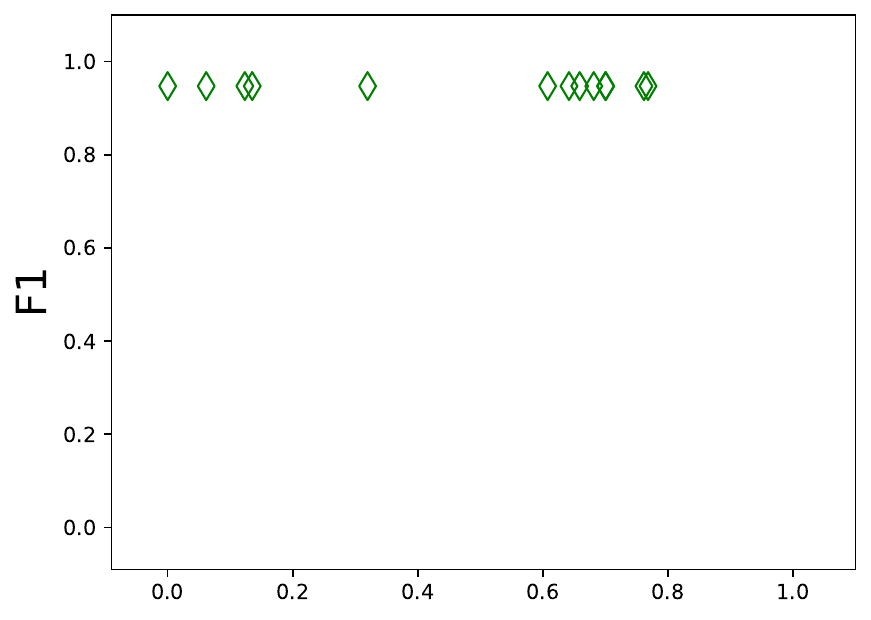}}
    &  \raisebox{-0.5\height}{\includegraphics[width=0.29\textwidth] {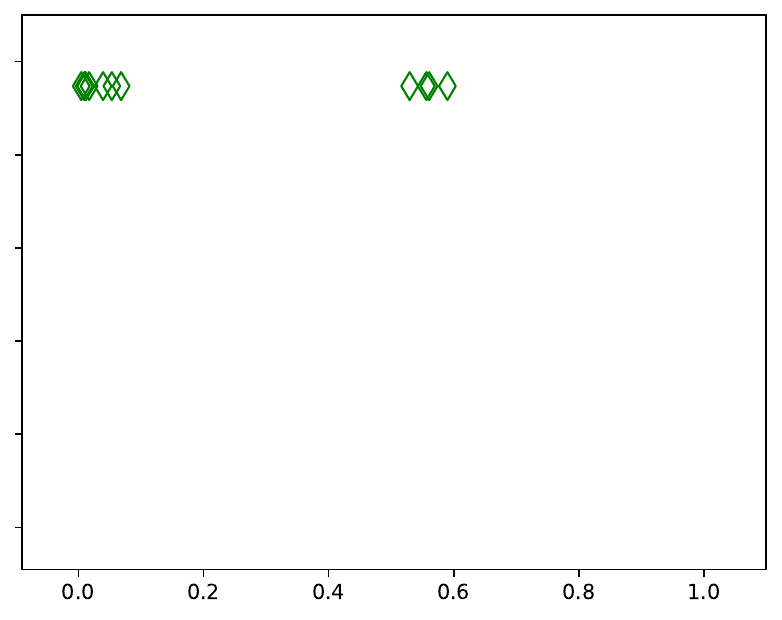}}
    &  \raisebox{-0.5\height}{\includegraphics[width=0.29\textwidth] {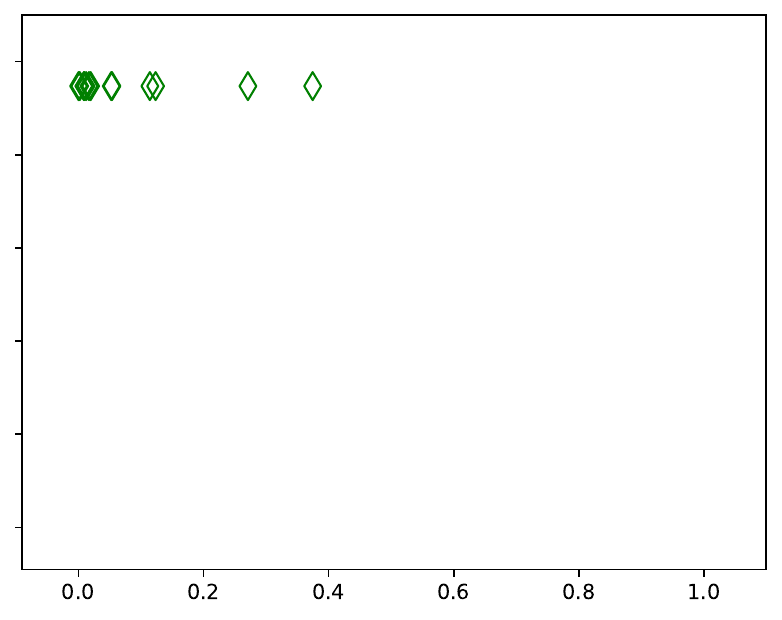}}
    \rotatebox[origin=c]{270}{\textbf{LogAnomaly}} \\

    & \raisebox{-0.5\height}{\includegraphics[width=0.30\textwidth] {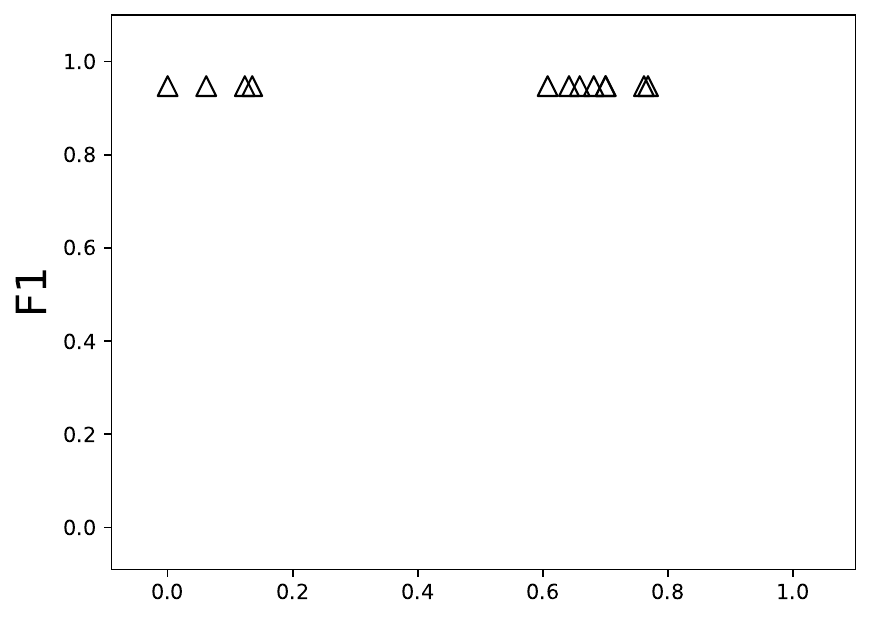}}
    & \raisebox{-0.5\height}{\includegraphics[width=0.29\textwidth] {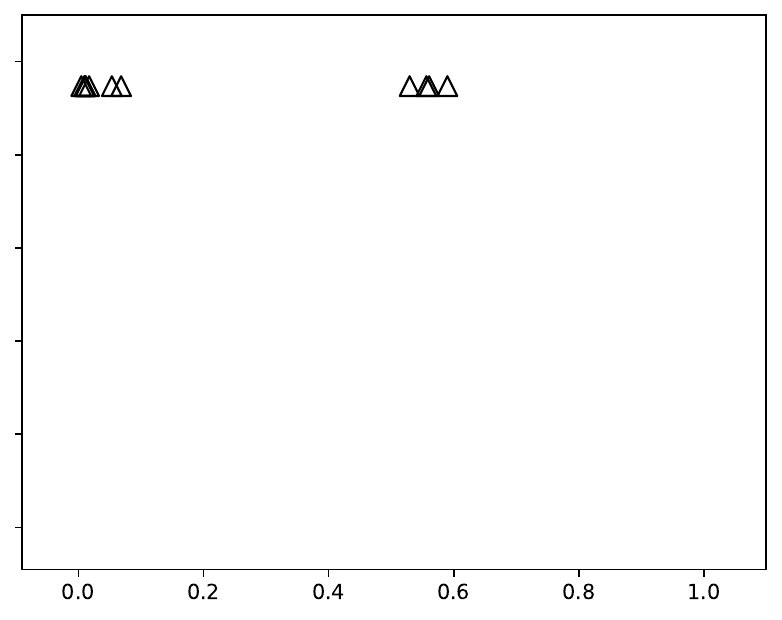}}
    & \raisebox{-0.5\height}{\includegraphics[width=0.29\textwidth] {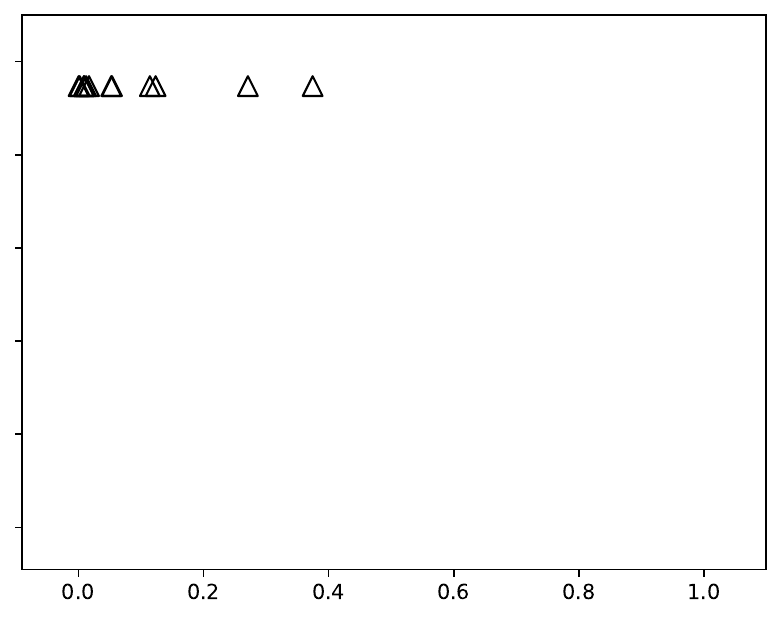}}
    \rotatebox[origin=c]{270}{\textbf{LogRobust}} \\

    & \raisebox{-0.5\height}{\includegraphics[width=0.30\textwidth] {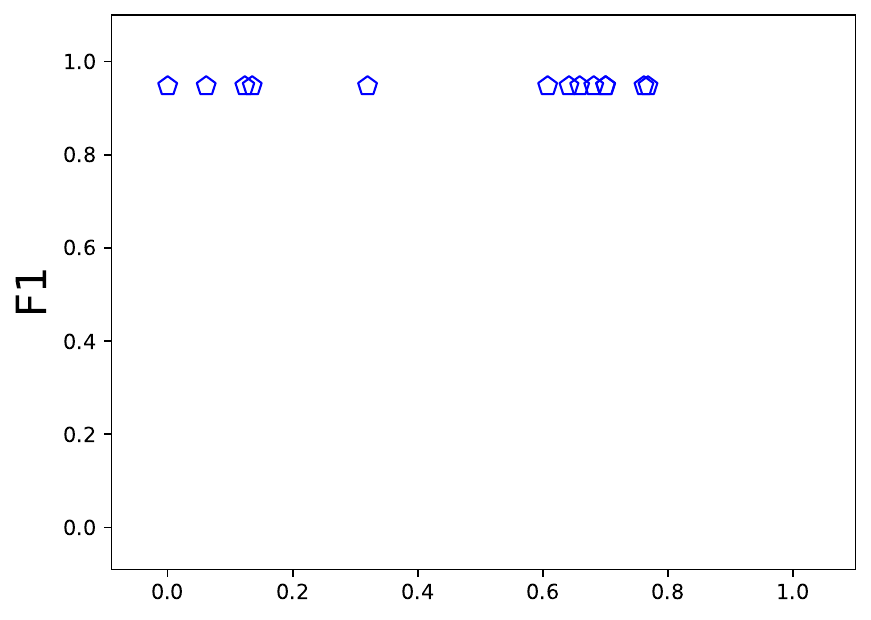}}
    & \raisebox{-0.5\height}{\includegraphics[width=0.29\textwidth] {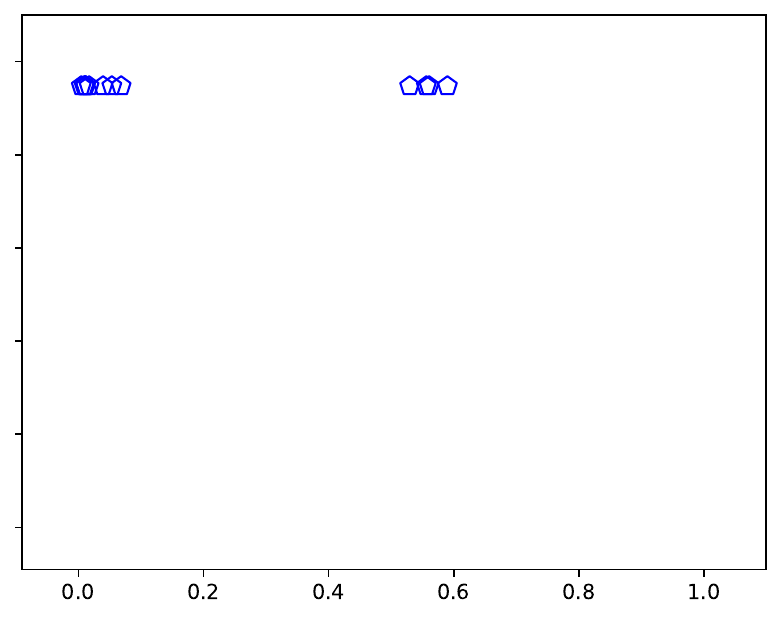}}
    & \raisebox{-0.5\height}{\includegraphics[width=0.29\textwidth] {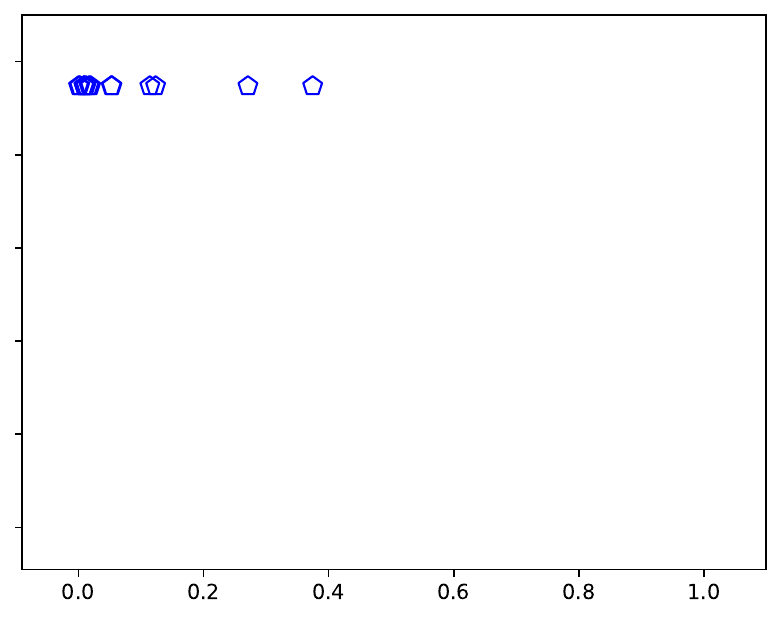}}
    \rotatebox[origin=c]{270}{\textbf{CNN}} \\

    & \raisebox{-0.5\height}{\includegraphics[width=0.30\textwidth] {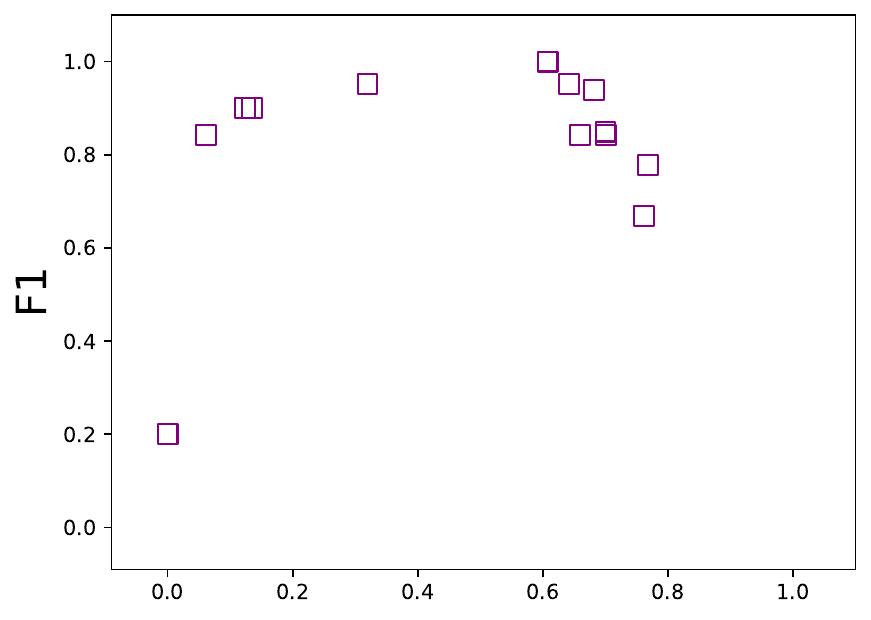}}
    & \raisebox{-0.5\height}{\includegraphics[width=0.29\textwidth] {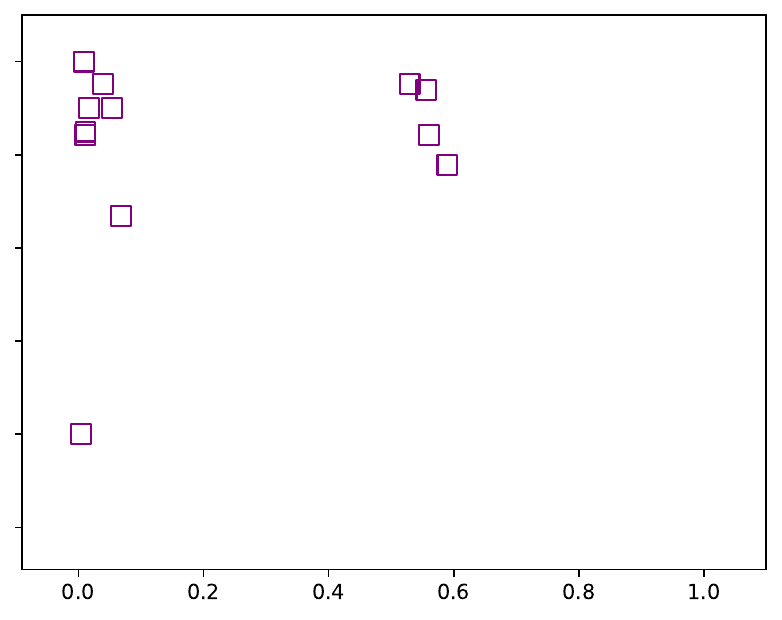}}
    & \raisebox{-0.5\height}{\includegraphics[width=0.29\textwidth] {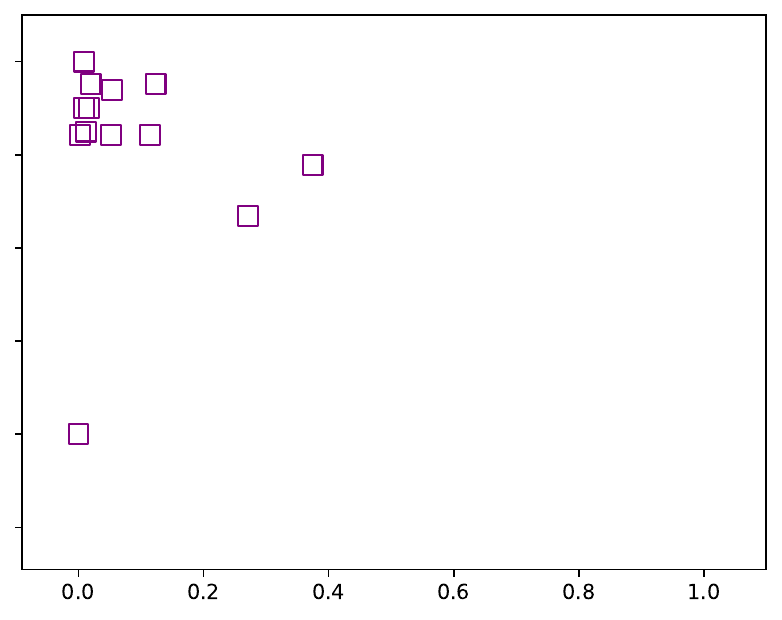}}
    \rotatebox[origin=c]{270}{\textbf{PLELog}} \\

     & \raisebox{-0.5\height}{\includegraphics[width=0.30\textwidth] {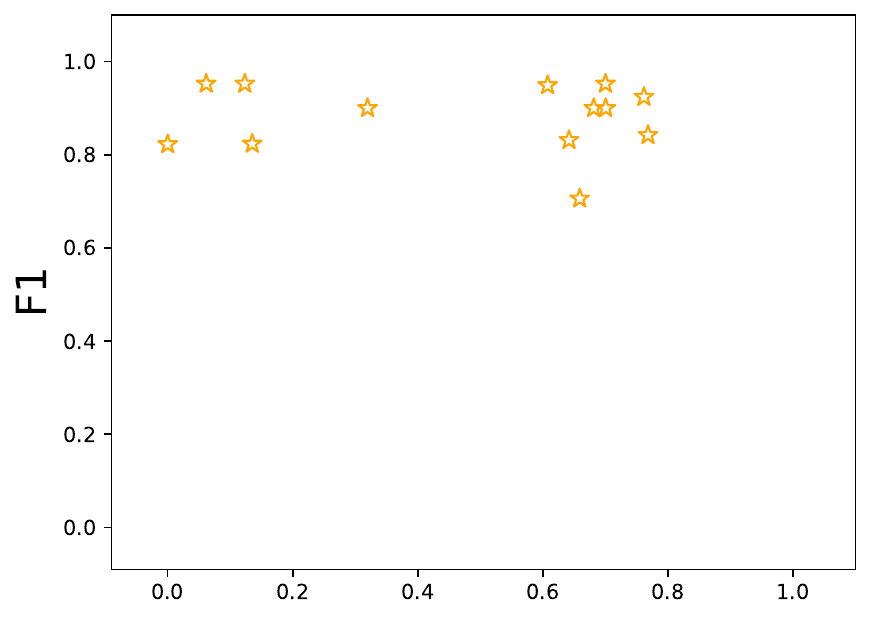}}
    & \raisebox{-0.5\height}{\includegraphics[width=0.29\textwidth] {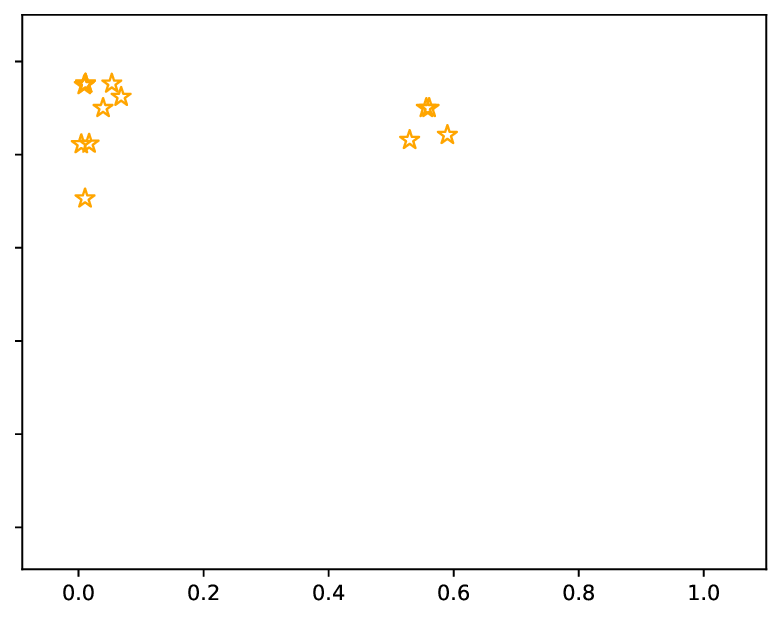}}
    & \raisebox{-0.5\height}{\includegraphics[width=0.29\textwidth] {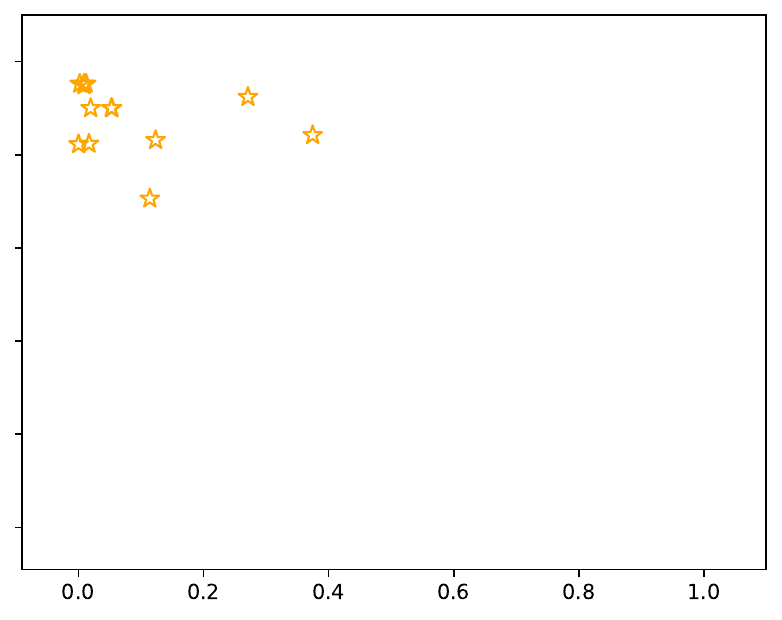}}
    \rotatebox[origin=c]{270}{\textbf{SVM}} \\

     & \raisebox{-0.5\height}{\includegraphics[width=0.30\textwidth] {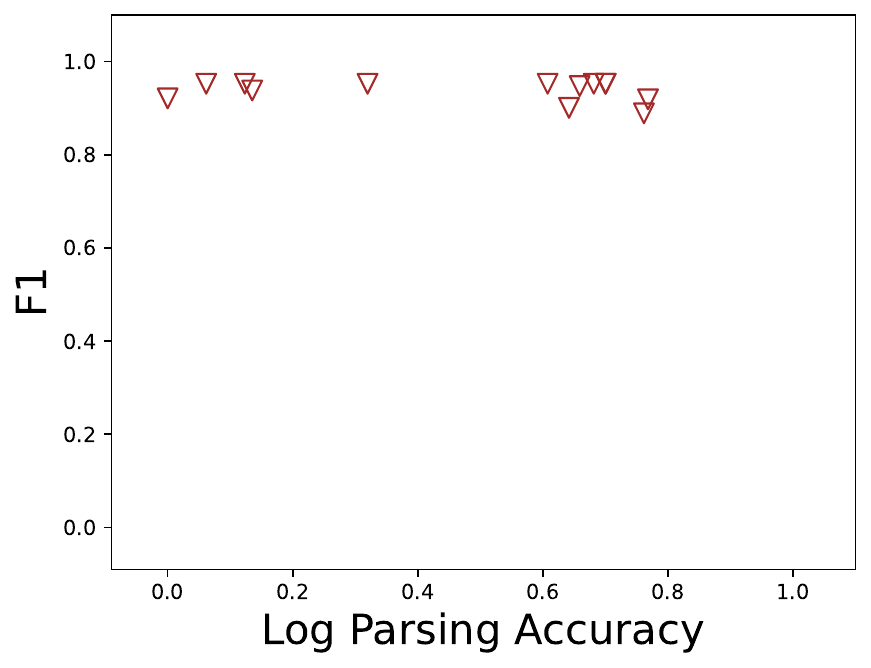}}
    & \raisebox{-0.5\height}{\includegraphics[width=0.29\textwidth] {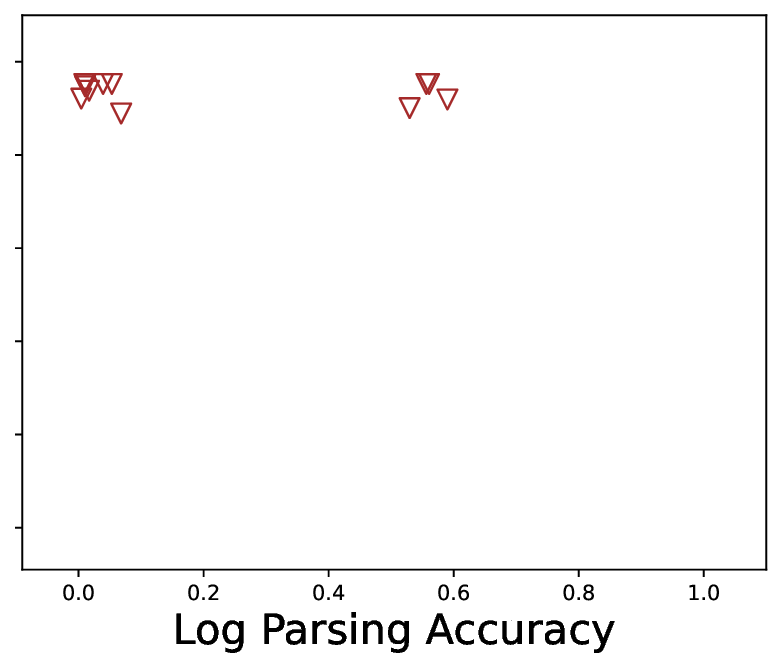}}
    & \raisebox{-0.5\height}{\includegraphics[width=0.29\textwidth] {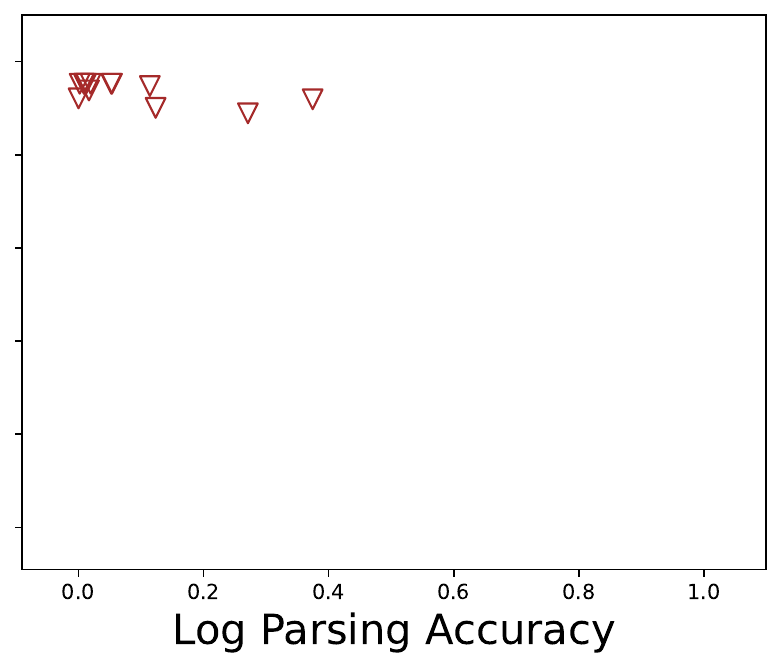}}
    \rotatebox[origin=c]{270}{\textbf{RF}} \
    
  \end{tabularx}

\caption{Relationship between TI accuracy and AD accuracy (Hadoop)
}
\label{fig:rq1-hadoop-results}
\end{figure}

For all tuples $\langle \mathit{GA, PA, FTA, PR, RE, F1} \rangle$ we collected for HDFS and Hadoop,
\figurename~\ref{fig:rq1-hdfs-results} and \figurename~\ref{fig:rq1-hadoop-results} show the relationship between
$\langle \mathit{GA, PA, FTA} \rangle$ (x-axis) and $F1$ (y-axis) for
HDFS and Hadoop, respectively, in the form of a scatter plot. To additionally distinguish the
main
results for different anomaly detection techniques, we used different shapes and
colors: {\protect\tikz \protect\draw[thick, draw=red, fill=white]
	plot[mark=*, mark options={scale=1.2}]
	(0,0);} = DeepLog,
	{\protect\tikz \protect\draw[thick, draw=teal, fill=white]
	plot[mark=diamond, mark options={scale=1.2}]
	(0,0);} = LogAnomaly,
	{\protect\tikz \protect\draw[thick, draw=black, fill=white]
	plot[mark=triangle*, mark options={scale=1.2}]
	(0,0);} = LogRobust,
	{\protect\tikz \protect\draw[thick, draw=blue, fill=white]
	plot[mark=pentagon*, mark options={scale=1.2}]
	(0,0);} = CNN,
	{\protect\tikz \protect\draw[thick, draw=purple, fill=white]
	plot[mark=square, mark options={scale=1.2}]
	(0,0);} = PLELog, 
	{\protect\tikz \protect\draw[thick, draw=orange, fill=white]
        plot[mark=star, mark options={scale=1.2, fill=white}]
        (0,0);}= \rev{SVM}, and
	{\protect\tikz \protect\draw[thick, draw=brown, fill=white]
	plot[mark=triangle*, mark options={scale=1.2, rotate=180}]
	(0,0);} = \rev{RF}. 
For example, the top left subfigure in \figurename~\ref{fig:rq1-hdfs-results} shows 13 data points
where 13 log parsing techniques are used in combination with DeepLog. All the raw data are available in the replication package on Figshare~\cite{figshare}.

Table~\ref{table:spearman_correlation_coefficient} additionally shows
the values of the Spearman's
rank correlation coefficient $\sigma \langle \mathit{X, Y} \rangle$ between $X= \langle \mathit{GA,
PA, FTA} \rangle$ and $Y= F1$
for each pair of anomaly detection technique and dataset. 
The value of $\sigma \langle \mathit{X, Y} \rangle$,
ranging between $-1$ and $+1$, is an indication of the strength of the monotonic (not
necessarily linear) relationship between $X$ and $Y$; when
$\sigma\langle \mathit{X, Y} \rangle \ge +0.7$ 
(or $\sigma\langle \mathit{X, Y} \rangle \le -0.7$), there is a
\textit{strong} positive (or negative) correlation between $X$ and $Y$~\cite{ali2022spearman}. Note
that, on the Hadoop dataset, $\sigma\langle \mathit{X, Y} \rangle$
could not be computed for DeepLog, LogAnomaly,
LogRobust, and CNN since the F1 score does not vary at all with
$\langle \mathit{GA, PA, FTA} \rangle$, indicating no relationship.

\begin{table*}
\small
\centering
\caption{Spearman correlation coefficients between log parsing accuracy (GA, PA, and FTA)
and anomaly detection accuracy (F1 score)}
\label{table:spearman_correlation_coefficient}
\begin{filecontents*}{data/correlation_results.csv}
AD_metric,HDFS_GA,HDFS_PA,HDFS_FTA,Hadoop_GA,Hadoop_PA,Hadoop_FTA
DeepLog,-0.165748386,0.25944476,0.198070731,,,
LogAnomaly,0.430945804,0.454725763,0.527258706,,,
LogRobust,0.215769492,-0.134091291,-0.162026977,,,
CNN,0.27624731,0.262234489,0.195281002,,,
PLELog,0.171273332,0.655586222,0.627688936,-0.180061619,-0.002770179,-0.069254469
SVM,0.633090492,0.37056087,0.650469509,-0.011065272,0.05256004,-0.345789735
RF,0.118321596,0.204838281,-0.06258948,-0.303417523,-0.135815463,-0.569269067
\end{filecontents*}
\pgfplotstabletypeset[
	col sep=comma,
	empty cells with={-}, string replace={---}{\textemdash},
	every head row/.style={
		before row={\toprule
			\multicolumn{1}{c}{} & 
			\multicolumn{3}{c}{HDFS (reduced)} &
			\multicolumn{3}{c}{Hadoop}
			\\
			\cmidrule(r){2-4}
			\cmidrule(r){5-7}},
		after row=\midrule,
	},
	every last row/.style={after row=\bottomrule},
	every column/.style={fixed zerofill,precision=3},
	columns/AD_metric/.style={column type=l,column name=AD technique,string type},
	columns/HDFS_GA/.style={column type=r, column name=GA, precision=3},
	columns/HDFS_PA/.style={column type=r, column name=PA, precision=3},
	columns/HDFS_FTA/.style={column type=r, column name=FTA, precision=3, fixed},
	columns/Hadoop_GA/.style={column type=r, column name=GA, precision=3, fixed}, columns/Hadoop_PA/.style={column type=r, column name=PA, precision=3, fixed},
	columns/Hadoop_FTA/.style={column type=r, column name=FTA, precision=3, fixed},
]{data/correlation_results.csv}
 \end{table*}

Overall, \figurename~\ref{fig:rq1-hdfs-results}, \figurename~\ref{fig:rq1-hadoop-results}, and
Table~\ref{table:spearman_correlation_coefficient} clearly show that there is no
strong correlation between $\langle \mathit{GA, PA, FTA} \rangle$ and $F1$ in all the
cases where $\langle \mathit{GA, PA, FTA, PR,  RE, F1} \rangle$ tuples were successfully collected. 
For example, in \figurename~\ref{fig:rq1-hdfs-results}, LogAnomaly 
({\protect\tikz \protect\draw[thick, draw=teal, fill=white] 
plot[mark=diamond, mark options={scale=1.2}] (0,0);}) achieved an F1 score
ranging between 0.2 and 0.5 regardless of the GA score. This means that increasing log
parsing accuracy does not necessarily increase (or decrease) anomaly detection accuracy.
This is counter-intuitive since anomaly detection uses log parsing results, and
having ``better'' log parsing results is expected to increase anomaly detection
accuracy. However, this happens because even inaccurate log parsing results can
lead to accurate anomaly detection results, for reasons explained below.

To better understand the reason for the above results, 
let us consider the following two extreme cases separately:
\begin{enumerate}[(C1)]
\item The log parsing accuracy values for input logs are the same, 
but the resulting anomaly detection accuracy values are different 
(i.e., the data points located on the same vertical lines in 
\figurename~\ref{fig:rq1-hdfs-results} and \figurename~\ref{fig:rq1-hadoop-results}).
\item The log parsing accuracy values for input logs are different, 
but the resulting anomaly detection accuracy values are the same 
(i.e., the data points located on the same horizontal lines in 
\figurename~\ref{fig:rq1-hdfs-results} and \figurename~\ref{fig:rq1-hadoop-results}). 
\end{enumerate}

To identify the root cause of C1, we manually investigated several pairs of data points in 
\figurename~\ref{fig:rq1-hdfs-results} and \figurename~\ref{fig:rq1-hadoop-results}, 
such as two different HDFS log parsing results having almost the same log parsing accuracy value 
(GA scores of 0.37 and 0.40) but resulting in significantly different anomaly detection accuracy values 
(F1 scores of 0.73 and 0.10) for the same anomaly detection technique (DeepLog). 
It turned out that, although the log parsing accuracy values are similar, 
\rev{the sets of correctly parsed log messages are different.}\label{page:r1c9:e}
This happened because the log parsing accuracy metrics (GA, PA, and FTA) summarize 
the log parsing results based on an implicit assumption that 
all log messages (and templates) are equally important.
However, this assumption does not hold when it comes to anomaly detection, 
which must discriminate different log message templates to learn abnormal sequences of templates.
Therefore, this mismatch of assumptions between log parsing and anomaly detection leads to case C1.

As for case C2, similar to the above case, 
we manually investigated several pairs of data points in 
\figurename~\ref{fig:rq1-hdfs-results} and \figurename~\ref{fig:rq1-hadoop-results}, 
such as two different Hadoop log parsing results having significantly different log parsing 
accuracy values (GA scores of 0.12 and 0.77) but resulting in the same anomaly detection value 
(F1 score of 0.98) for the same anomaly detection technique (DeepLog). 
We found that anomaly detection techniques can distinguish between normal and abnormal patterns 
even when input log message templates are incorrect. 
To best explain this using a simplified example, let us consider a normal log
$l_n = \langle \mathit{m_1^n, m_2^n}, \dots \rangle$ and an abnormal log $l_a = \langle
\mathit{m_1^a, m_2^a}, \dots \rangle$, where $m_i^x$ indicates the $i$-th log message in
$l_x$ for $x\in \{n, a\}$. Using oracle templates, we can group the log messages
having the same template and represent $l_n$ and $l_a$ as groups; 
specifically, let $g_\mathit{orc}(l_x)$ be a sequence of
message group indices (i.e., the $i$-th element of $g_\mathit{orc}(l_x)$ is
the message group index of $m_i^x$). 
In this context, let us take two logs from
the Hadoop dataset as a concrete example where 
$g_\mathit{orc}(l_n) = \langle 1, 2, 3, 4, \dots \rangle$ and 
$g_\mathit{orc}(l_a) = \langle 5, 5, 5, 6, \dots \rangle$. 
When templates generated by LogMine are used to group
messages instead of oracle templates, the sequences of message group indices
change to $g_\mathit{LM}(l_n) = \langle 1, 2, 3, 3, \dots \rangle$
and $g_\mathit{LM}(l_a) = \langle 7, 8, 9, 10, \dots \rangle$. 
These are clearly different from $g_\mathit{orc}(l_n)$ and $g_\mathit{orc}(l_a)$, respectively; 
in particular, $m_3^n$ and $m_4^n$ are incorrectly grouped together in $g_\mathit{LM}(l_n)$ 
while $m_1^a$, $m_2^a$, and $m_3^a$ are incorrectly separated in $g_\mathit{LM}(l_a)$. 
The incorrect groupings of LogMine clearly reduce the GA score (as well as PA and TA
scores since incorrect groupings 
imply incorrect templates). However, even the incorrect $g_\mathit{LM}(l_n)$ and
$g_\mathit{LM}(l_a)$ are still different enough from each other for anomaly detection techniques to distinguish between normal and abnormal patterns.
This example not only shows why case C2 happened,
but also demonstrates the importance of \textit{distinguishability} in log parsing results for anomaly
detection; we will further investigate this aspect in RQ2.

Before we conclude RQ1, one might be curious to know why DeepLog, LogAnomaly, LogRobust,
and CNN result in the same anomaly detection accuracy value on the Hadoop
dataset (as shown in Figure~\ref{fig:rq1-hadoop-results} [GA-Hadoop] and
Table~\ref{table:spearman_correlation_coefficient}). This happens because (1) the
test set of Hadoop contains only 11 logs (1 normal and 10 abnormal logs, 
although the number of log messages is in the same order of magnitude as HDFS; 
see Table~\ref{table:datasets} for more details) and (2)
the four anomaly detection techniques classified all the 11 logs in the test set as
abnormal. We speculate that PLELog shows different results from the other anomaly
detection techniques because PLELog uses a very different deep learning model (i.e., an
attention-based GRU~\cite{cho-etal-2014}). Notice that, in all 
cases, the results still corroborate that log parsing accuracy and anomaly detection accuracy do not
have any strong relationship. 

We want to note that the log parsing accuracy results shown in 
\figurename~\ref{fig:rq1-hdfs-results} and \figurename~\ref{fig:rq1-hadoop-results} 
are inconsistent with the ones reported in previous studies~\cite{zhu2019tools,dai2020logram} 
since the latter only considered 2K log messages, randomly sampled from the original logs, 
to assess log parsing accuracy.

\begin{tcolorbox}
The answer to RQ1 is that there is no strong correlation between log
parsing accuracy and anomaly detection accuracy; increasing log parsing
accuracy does not necessarily increase anomaly detection accuracy, regardless of
the metric (GA, PA, or TA) used for measuring log parsing accuracy.
\end{tcolorbox}

 \subsection{RQ2: Log Parsing Distinguishability and Anomaly Detection Accuracy}\label{sec:eval-rq2-results}

\subsubsection{Distinguishability as a Binary Property}\label{sec:dist-binary}

\begin{table*}
  \caption{Impact of the distinguishability log parsing results on
    anomaly detection accuracy for the HDFS (reduced) dataset (DL-based anomaly detection techniques)}
  \label{table:rq2_hdfs}
  \begin{adjustbox}{width=\textwidth}
    \begin{filecontents*}{data/HDFS_RQ2.csv}
TI,AD_DeepLog_F1_dist,AD_DeepLog_F1_indist,DeepLog_delta,AD_LogAnomaly_F1_dist,AD_LogAnomaly_F1_indist,LogAnomaly_delta,AD_LogRobust_F1_dist,AD_LogRobust_F1_indist,LogRobust_delta,AD_CNN_F1_dist,AD_CNN_F1_indist,CNN_delta,AD_PLELog_F1_dist,AD_PLELog_F1_indist,PLELog_delta
AEL,0.747,0.5609,0.1861,0.5086,0.3201,0.1885,0.6628,0.4556,0.2072,0.7719,0.6615,0.1104,0.7602,0.0334,0.7268
Drain,0.7138,0.5233,0.1905,0.4993,0.4,0.0993,0.7032,0.4536,0.2496,0.7572,0.6821,0.0751,0.796,0.2857,0.5103
IPLoM,0.76,0.5898,0.1702,0.4808,0.2678,0.213,0.556,0.3801,0.1759,0.8099,0.588,0.2219,0.8485,0.0408,0.8077
LFA,0.803,0.6926,0.1104,0.6063,0.378,0.2283,0.3551,0.2992,0.0559,0.7554,0.5484,0.207,0.1,0,0.1
LenMa,0.8083,0.6248,0.1835,0.4838,0.2849,0.1989,0.659,0.4357,0.2233,0.8135,0.6067,0.2068,0.6814,0.2706,0.4108
LogCluster,0.2631,0.0974,0.1657,0.3803,0.2427,0.1376,0.5416,0.3002,0.2414,0.4975,0.306,0.1915,0.4255,0.3172,0.1083
LogMine,0.7316,0.5521,0.1795,0.4533,0.363,0.0903,0.5535,0.329,0.2245,0.7915,0.6122,0.1793,0.8169,0.4387,0.3782
Logram,0.2018,0.0247,0.1771,0.2902,0.1426,0.1476,0.6962,0.4604,0.2358,0.6989,0.5225,0.1764,0.7868,0.034,0.7528
MoLFI,0.7941,0.6299,0.1642,0.4265,0.2822,0.1443,0.5645,0.3188,0.2457,0.7806,0.6211,0.1595,0.1722,0.1089,0.0633
SHISO,0.7777,0.6289,0.1488,0.5437,0.2382,0.3055,0.6791,0.4464,0.2327,0.7961,0.5894,0.2067,0.8389,0.3406,0.4983
SLCT,0.7429,0.5698,0.1731,0.2683,0.1603,0.108,0.3944,0.244,0.1504,0.7429,0.6065,0.1364,0.7245,0.5338,0.1907
Spell,0.7648,0.5983,0.1665,0.2886,0.176,0.1126,0.4007,0.2408,0.1599,0.805,0.6156,0.1894,0.6653,0.3041,0.3612
Average,0.675675,0.507708333,0.167966667,0.435808333,0.271316667,0.164491667,0.563841667,0.36365,0.200191667,0.7517,0.58,0.1717,0.634683333,0.22565,0.409033333
\end{filecontents*}
\pgfplotstabletypeset[
	col sep=comma,
	fixed, fixed zerofill,
	empty cells with={-}, precision=3,
	every head row/.style={
		before row={
	   		\toprule
			\multicolumn{1}{c}{} & 
			\multicolumn{3}{c}{DeepLog (F1)} &
			\multicolumn{3}{c}{LogAnomaly (F1)} &
			\multicolumn{3}{c}{LogRobust (F1)} &
			\multicolumn{3}{c}{CNN (F1)} &
			\multicolumn{3}{c}{PLELog (F1)} 
			\\
			\cmidrule(r){2-4}
			\cmidrule(r){5-7}
			\cmidrule(r){8-10}
			\cmidrule(r){11-13}
			\cmidrule(r){14-16}
		},
		after row=\midrule,
	},
	every last row/.style={
		before row=\midrule,
		after row=\bottomrule,
	},
	columns/TI/.style={column type=l,column name=Log Parser,string type},
	columns/AD_DeepLog_F1_dist/.style={column type=r,column name=$R_\mathit{dst}$},
	columns/AD_DeepLog_F1_indist/.style={column type=r, column name=$R_\mathit{ind}$},
	columns/DeepLog_delta/.style={column type=r, column name=$\Delta$},
	columns/AD_LogAnomaly_F1_dist/.style={column type=r,column name=$R_\mathit{dst}$},
	columns/AD_LogAnomaly_F1_indist/.style={column type=r, column name=$R_\mathit{ind}$},
	columns/LogAnomaly_delta/.style={column type=r, column name=$\Delta$},
	columns/AD_LogRobust_F1_dist/.style={column type=r,column name=$R_\mathit{dst}$},
	columns/AD_LogRobust_F1_indist/.style={column type=r, column name=$R_\mathit{ind}$},
	columns/LogRobust_delta/.style={column type=r, column name=$\Delta$},
	columns/AD_CNN_F1_dist/.style={column type=r,column name=$R_\mathit{dst}$},
	columns/AD_CNN_F1_indist/.style={column type=r,column name=$R_\mathit{ind}$},
	columns/CNN_delta/.style={column type=r, column name=$\Delta$},
	columns/AD_PLELog_F1_dist/.style={column type=r,column name=$R_\mathit{dst}$},
	columns/AD_PLELog_F1_indist/.style={column type=r,column name=$R_\mathit{ind}$},
	columns/PLELog_delta/.style={column type=r, column name=$\Delta$}
]{data/HDFS_RQ2.csv}
     \end{adjustbox}
  \end{table*}
  
  \begin{table*}
  \centering
  \caption{\rev{Impact of the distinguishability of log parsing results on
    anomaly detection accuracy for the HDFS (reduced) dataset (ML-based anomaly detection techniques)}}
  \label{table:rq2_hdfs_ml}
\begin{filecontents*}{data/HDFS_RQ2_ML.csv}
TI,AD_SVM_F1_dist,AD_SVM_F1_indist,SVM_delta,AD_RF_F1_dist,AD_RF_F1_indist,RF_delta
AEL,0.9677,0.9277,0.04,0.9471,0.6785,0.2686
Drain,0.9677,0.9277,0.04,0.954,0.6785,0.2755
IPLoM,0.9677,0.9277,0.04,0.9609,0.6785,0.2824
LFA,0.9677,0.9277,0.04,0.9471,0.6785,0.2686
LenMa,0.9677,0.9277,0.04,0.9471,0.6785,0.2686
LogCluster,0,0,0,0.2222,0,0.2222
LogMine,0.9471,0.9277,0.0194,0.9471,0.7822,0.1649
Logram,0.8676,0.7079,0.1597,0.9677,0.6015,0.3662
MoLFI,0.9677,0.9277,0.04,0.9586,0.6785,0.2801
SHISO,0.9609,0.9277,0.0332,0.9402,0.6785,0.2617
SLCT,0.9677,0.9277,0.04,0.9402,0.6785,0.2617
Spell,0.9677,0.9277,0.04,0.9471,0.6785,0.2686
Average,0.876433333,0.832075,0.044358333,0.889941667,0.624183333,0.265758333
\end{filecontents*}
\pgfplotstabletypeset[
	col sep=comma,
	fixed, fixed zerofill,
	empty cells with={-}, precision=3,
	every head row/.style={
		before row={
	   		\toprule
			\multicolumn{1}{c}{} & 
			\multicolumn{3}{c}{SVM (F1)} &
			\multicolumn{3}{c}{RF (F1)} 
			\\
			\cmidrule(r){2-4}
			\cmidrule(r){5-7}
		},
		after row=\midrule,
	},
	every last row/.style={
		before row=\midrule,
		after row=\bottomrule,
	},
	columns/TI/.style={column type=l,column name=Log Parser,string type},
		columns/AD_SVM_F1_dist/.style={column type=r,column name=$R_\mathit{dst}$},
	columns/AD_SVM_F1_indist/.style={column type=r,column name=$R_\mathit{ind}$},
	columns/SVM_delta/.style={column type=r, column name=$\Delta$},
		columns/AD_RF_F1_dist/.style={column type=r,column name=$R_\mathit{dst}$},
	columns/AD_RF_F1_indist/.style={column type=r,column name=$R_\mathit{ind}$},
	columns/RF_delta/.style={column type=r, column name=$\Delta$},
]{data/HDFS_RQ2_ML.csv}
 \end{table*}
  
  \begin{table*}
  \caption{Impact of the distinguishability of log parsing results on
    anomaly detection accuracy for the Hadoop dataset (DL-based anomaly detection techniques)}
  \label{table:rq2_hadoop}
  \begin{adjustbox}{width=\textwidth}
    \begin{filecontents*}{data/Hadoop_RQ2.csv}
TI,AD_DeepLog_F1_dist,AD_DeepLog_F1_indist,DeepLog_delta,AD_LogAnomaly_F1_dist,AD_LogAnomaly_F1_indist,LogAnomaly_delta,AD_LogRobust_F1_dist,AD_LogRobust_F1_indist,LogRobust_delta,AD_CNN_F1_dist,AD_CNN_F1_indist,CNN_delta,AD_PLELog_F1_dist,AD_PLELog_F1_indist,PLELog_delta
AEL,0.9474,0.8471,0.1003,0.9474,0.8471,0.1003,0.9474,0.8471,0.1003,0.9474,0.8471,0.1003,0.7845,0.5066,0.2779
Drain,0.9474,0.8471,0.1003,0.9474,0.8471,0.1003,0.9474,0.8471,0.1003,0.9474,0.8471,0.1003,0.9,0,0.9
IPLoM,0.9474,0.8471,0.1003,0.9474,0.8471,0.1003,0.9474,0.8471,0.1003,0.9474,0.8471,0.1003,0.8482,0,0.8482
LFA,0.9474,0.8471,0.1003,0.9474,0.8471,0.1003,0.9474,0.8471,0.1003,0.9474,0.8471,0.1003,0.8884,0.7989,0.0895
LenMa,0.9474,0.8471,0.1003,0.9474,0.8471,0.1003,0.9474,0.8471,0.1003,0.9474,0.8471,0.1003,0.9,0.5048,0.3952
LogCluster,0.9474,0.8471,0.1003,0.9474,0.8471,0.1003,0.9474,0.8471,0.1003,0.9474,0.8471,0.1003,0.9524,0.18,0.7724
LogMine,0.9474,0.8471,0.1003,0.9474,0.8471,0.1003,0.9474,0.8471,0.1003,0.9474,0.8471,0.1003,0.8482,0.5303,0.3179
Logram,0.9474,0.8471,0.1003,0.9474,0.8471,0.1003,0.9474,0.8471,0.1003,0.9474,0.8471,0.1003,0.8421,0.7989,0.0432
MoLFI,0.9474,0.8471,0.1003,0.9474,0.8471,0.1003,0.9474,0.8471,0.1003,0.9474,0.8471,0.1003,0.9,0.7393,0.1607
SHISO,0.9474,0.8471,0.1003,0.9474,0.8471,0.1003,0.9474,0.8471,0.1003,0.9474,0.8471,0.1003,0.9,0,0.9
SLCT,0.9474,0.8471,0.1003,0.9474,0.8471,0.1003,0.9474,0.8471,0.1003,0.9474,0.8471,0.1003,0.8421,0,0.8421
Spell,0.9474,0.8471,0.1003,0.9474,0.8471,0.1003,0.9474,0.8471,0.1003,0.9474,0.8471,0.1003,0.7029,0.1883,0.5146
Average,0.9474,0.8471,0.1003,0.9474,0.8471,0.1003,0.9474,0.8471,0.1003,0.9474,0.8471,0.1003,0.859066667,0.353925,0.505141667
\end{filecontents*}
\pgfplotstabletypeset[
	col sep=comma,
	fixed zerofill, fixed,
	precision=3,
	every head row/.style={
		before row={
	   		\toprule
			\multicolumn{1}{c}{} & 
			\multicolumn{3}{c}{DeepLog (F1)} &
			\multicolumn{3}{c}{LogAnomaly (F1)} &
			\multicolumn{3}{c}{LogRobust (F1)} &
			\multicolumn{3}{c}{CNN (F1)} &
			\multicolumn{3}{c}{PLELog (F1)} 
			\\
			\cmidrule(r){2-4}
			\cmidrule(r){5-7}
			\cmidrule(r){8-10}
			\cmidrule(r){11-13}
			\cmidrule(r){14-16}
		},
		after row=\midrule,
	},
	every last row/.style={
		before row=\midrule,
		after row=\bottomrule,
	},
	columns/TI/.style={column type=l,column name=Log Parser, string type },
	columns/AD_DeepLog_F1_dist/.style={column type=r,column name=$R_\mathit{dst}$},
	columns/AD_DeepLog_F1_indist/.style={column type=r, column name=$R_\mathit{ind}$},
	columns/DeepLog_delta/.style={column type=r, column name=$\Delta$},
	columns/AD_LogAnomaly_F1_dist/.style={column type=r,column name=$R_\mathit{dst}$},
	columns/AD_LogAnomaly_F1_indist/.style={column type=r, column name=$R_\mathit{ind}$},
	columns/LogAnomaly_delta/.style={column type=r, column name=$\Delta$},
	columns/AD_LogRobust_F1_dist/.style={column type=r,column name=$R_\mathit{dst}$},
	columns/AD_LogRobust_F1_indist/.style={column type=r, column name=$R_\mathit{ind}$},
	columns/LogRobust_delta/.style={column type=r, column name=$\Delta$},
	columns/AD_CNN_F1_dist/.style={column type=r,column name=$R_\mathit{dst}$},
	columns/AD_CNN_F1_indist/.style={column type=r,column name=$R_\mathit{ind}$},
	columns/CNN_delta/.style={column type=r, column name=$\Delta$},
	columns/AD_PLELog_F1_dist/.style={column type=r,column name=$R_\mathit{dst}$},
	columns/AD_PLELog_F1_indist/.style={column type=r,column name=$R_\mathit{ind}$},
	columns/PLELog_delta/.style={column type=r, column name=$\Delta$}
]{data/Hadoop_RQ2.csv}
       \end{adjustbox}
  \end{table*}
  
  \begin{table*}
  \centering
  \caption{\rev{Impact of the distinguishability of log parsing results on
    anomaly detection accuracy for the Hadoop dataset (ML-based anomaly detection techniques)}}
  \label{table:rq2_hadoop_ml}
\begin{filecontents*}{data/Hadoop_RQ2_ML.csv}
TI,AD_SVM_F1_dist,AD_SVM_F1_indist,SVM_delta,AD_RF_F1_dist,AD_RF_F1_indist,RF_delta
AEL,0.96,0.9123,0.0477,0.9524,0.718,0.2344
Drain,0.9524,0.9123,0.0401,0.9095,0.8622,0.0473
IPLoM,0.9491,0.9123,0.0368,0.8912,0.8622,0.029
LFA,0.9397,0.9123,0.0274,0.9006,0.8622,0.0384
LenMa,0.9524,0.9123,0.0401,0.9524,0.718,0.2344
LogCluster,0.9524,0.9123,0.0401,0.9524,0.718,0.2344
LogMine,0.96,0.9123,0.0477,0.9524,0.718,0.2344
Logram,0.9524,0.85213,0.10027,0.7976,0.6428,0.1548
MoLFI,0.949,0.9123,0.0367,0.9524,0.718,0.2344
SHISO,0.9235,0.9123,0.0112,0.9357,0.8622,0.0735
SLCT,0.9366,0.9123,0.0243,0.9357,0.8622,0.0735
Spell,0.96,0.9123,0.0477,0.9524,0.718,0.2344
Average,0.948958333,0.907285833,0.0416725,0.923725,0.771816667,0.151908333
\end{filecontents*}
\pgfplotstabletypeset[
	col sep=comma,
	fixed zerofill, fixed,
	precision=3,
	every head row/.style={
		before row={
	   		\toprule
			\multicolumn{1}{c}{} & 
			\multicolumn{3}{c}{SVM (F1)} &
			\multicolumn{3}{c}{RF (F1)} 
			\\
			\cmidrule(r){2-4}
			\cmidrule(r){5-7}
		},
		after row=\midrule,
	},
	every last row/.style={
		before row=\midrule,
		after row=\bottomrule,
	},
	columns/TI/.style={column type=l,column name=Log Parser, string type },
			columns/AD_SVM_F1_dist/.style={column type=r,column name=$R_\mathit{dst}$},
	columns/AD_SVM_F1_indist/.style={column type=r,column name=$R_\mathit{ind}$},
	columns/SVM_delta/.style={column type=r, column name=$\Delta$},
		columns/AD_RF_F1_dist/.style={column type=r,column name=$R_\mathit{dst}$},
	columns/AD_RF_F1_indist/.style={column type=r,column name=$R_\mathit{ind}$},
	columns/RF_delta/.style={column type=r, column name=$\Delta$},
]{data/Hadoop_RQ2_ML.csv}
 \end{table*}

\rev{Tables~\ref{table:rq2_hdfs} and~\ref{table:rq2_hdfs_ml} show} the anomaly detection 
accuracy values (F1 scores) when different log parsing techniques (rows) and anomaly detection 
techniques (columns) are used together on the HDFS (reduced) dataset; under each of the 
anomaly detection technique columns, sub-columns $R_\mathit{dst}$ and 
$R_\mathit{ind}$ indicate the F1 scores for distinguishable and 
indistinguishable log parsing results, respectively, and $\Delta$ indicates the 
difference between $R_\mathit{dst}$ and $R_\mathit{ind}$. For example, if we 
choose AEL for log parsing and DeepLog for anomaly detection, the F1 score 
decreases from $0.747$ to $0.561$  when 
$R_\mathit{ind}$ is used instead of $R_\mathit{dst}$. 
The same structure applies to \rev{Tables~\ref{table:rq2_hadoop} and~\ref{table:rq2_hadoop_ml}}, 
which show the results on the Hadoop dataset.
In Table~\ref{table:rq2_hadoop}, except for PLELog, SVM, and RF, the values for all anomaly detection techniques 
are identical due to the reasons explained in the last 
paragraph of Section~\ref{sec:eval-rq1-results}.
\rev{We do not provide results for the OpenStack dataset due to the reasons mentioned in Section~\ref{sec:eval-rq1-results}.}

In all cases, $\Delta$ is non-negative, 
ranging from \rev{$0$ (LogCluster-SVM on the HDFS dataset) to 
$0.9$ (Drain/SHISO-PLELog on the Hadoop dataset)}. 
This means that the anomaly detection accuracy decreases up to 90 
percentage points (pp) when $R_\mathit{ind}$ is used instead of 
$R_\mathit{dst}$. To see if the differences between $R_\mathit{dst}$ and 
$R_\mathit{ind}$ are significant, we applied the non-parametric Wilcoxon signed rank 
test~\cite{wilcoxon1992individual} for 
paired samples to the F1 scores of $R_\mathit{dst}$ and $R_\mathit{ind}$, for 
each of the \rev{seven} anomaly detection techniques and the two datasets. The results show 
that, for all the anomaly detection techniques and datasets, the differences 
between $R_\mathit{dst}$ and $R_\mathit{ind}$ are significant
($p\text{-value} < 0.005$) in terms of anomaly detection accuracy.

Considering the definition of distinguishability for log parsing results, it 
is intuitive that indistinguishable log parsing results should lead to lower anomaly 
detection accuracy. However, it is surprising that this decrease in accuracy is, 
in some cases, rather limited, e.g., only \rev{$0.011$ for SHISO on the Hadoop dataset 
when SVM} is used for log parsing.
This happens because an indistinguishable log parsing 
result may only have a few logs that are indistinguishable in terms of normal and 
abnormal behavior. Recall that we did not explicitly control the number of 
indistinguishable logs since we aimed to minimize the difference between 
distinguishable and indistinguishable versions of each log parsing result as 
described in Section~\ref{sec:eval-rq2-method}. Nevertheless, the results shown 
in Tables~\ref{table:rq2_hdfs} and \ref{table:rq2_hadoop} are sufficient to confirm 
the strong impact of distinguishability in log parsing results on anomaly detection 
accuracy.

\begin{tcolorbox}
The answer to RQ2 is that the impact of the distinguishability 
of log parsing results on anomaly detection accuracy is significant 
for all anomaly detection techniques. 
\end{tcolorbox}

\subsubsection{\rev{Degree of Distinguishability}}\label{sec:dist-degree-results}

\rev{As explained in Section~\ref{sec:dist-degree}, let us consider the degree of 
distinguishability of the log parsing results generated for RQ1 
(without considering the artificially generated pairs of $R_{dst}$ and $R_{ind}$.)
We focus on the HDFS dataset for this analysis since we know from the RQ1 results that
(1) none of the anomaly detection techniques detected abnormal logs in the OpenStack dataset, and
(2) most of the anomaly detection techniques have achieved the same accuracy on the Hadoop dataset.}
\minrev{Nevertheless, to avoid drawing conclusions based on a single dataset, we also include another dataset, BGL, in this analysis.
Although it was excluded from the previous analyses due to the unavailability of
source code (which is essential to measure log parsing accuracy), it can be used to investigate the relationship between the degree of distinguishability and anomaly detection accuracy.
To use the BGL dataset, we first reduced it following the same methodology we used for the other datasets (see Section~\ref{sec:datasets}).
Since the dataset has only one extremely long normal log, we created log sequences using a sliding window with a window size of 10, following existing studies~\cite{yang2021semi, le2022log}.
We then labelled each log sequence as normal or abnormal as follows:
If a log sequence contains at least one abnormal log message, it is considered abnormal; otherwise, it is considered normal.
In total, we used \num{275306} normal and \num{16413} abnormal log sequences from the BGL dataset.}

\begin{figure}
  \centering
  
\begin{tabularx}{\textwidth}{
          l X  @{\hspace{3pt}}  X@{\hspace{1pt}}  l 
    }

      & \raisebox{-0.5\height}{\includegraphics[width=0.29\textwidth] {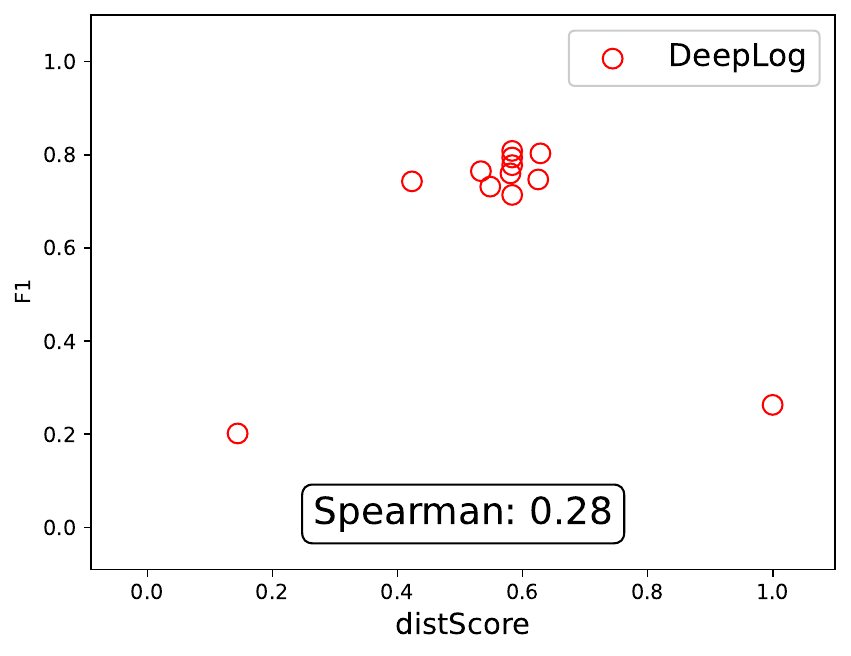}}
      & \raisebox{-0.5\height}{\includegraphics[width=0.29\textwidth] {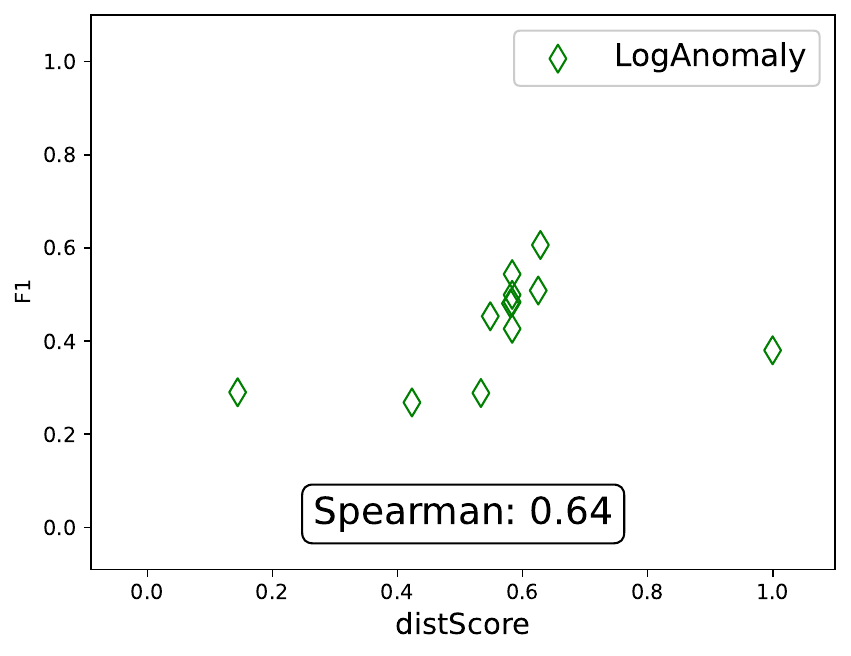}}
       & \raisebox{-0.5\height}{\includegraphics[width=0.29\textwidth] {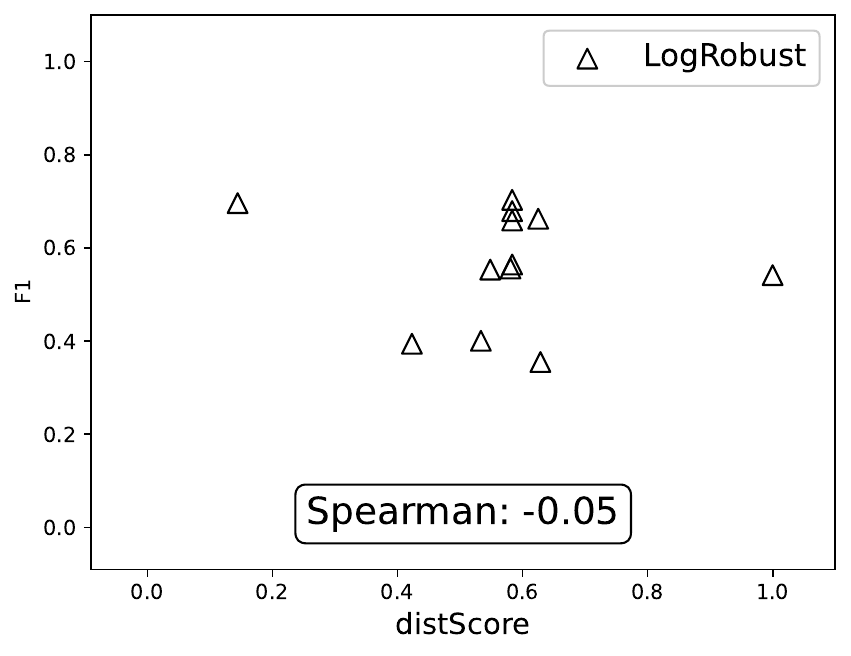}} \\
      
      &  \raisebox{-0.5\height}{\includegraphics[width=0.29\textwidth] {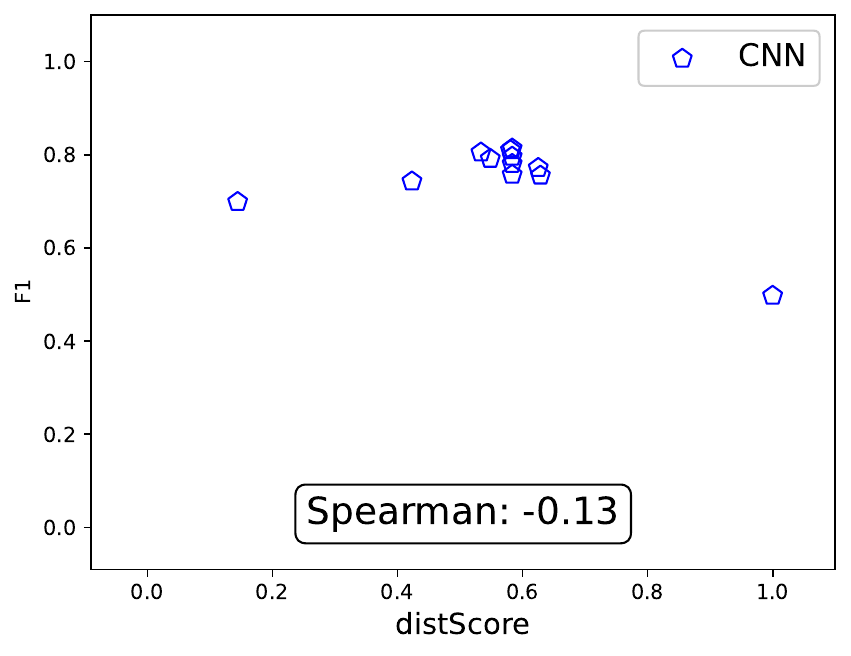}}
      &  \raisebox{-0.5\height}{\includegraphics[width=0.29\textwidth] {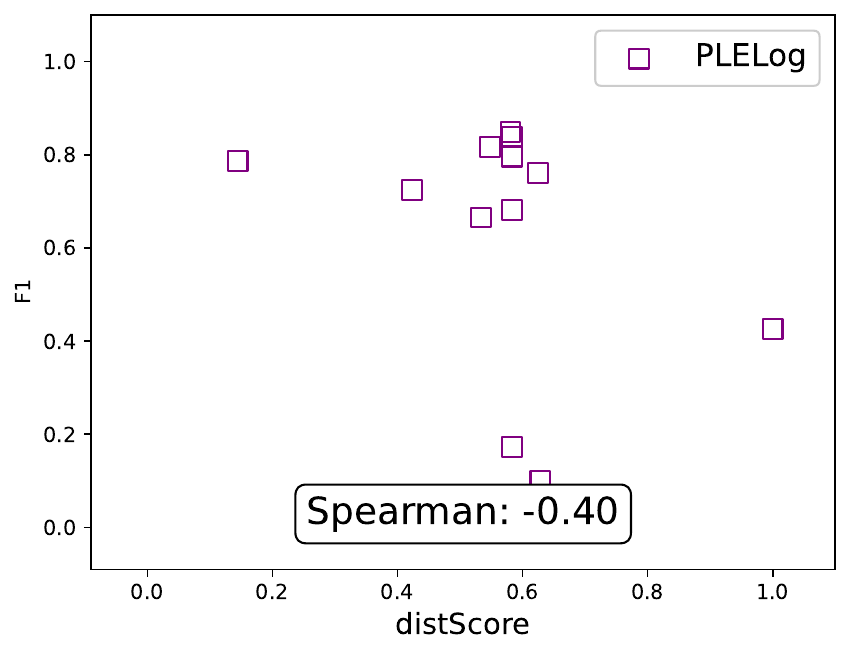}}
      &  \raisebox{-0.5\height}{\includegraphics[width=0.29\textwidth] {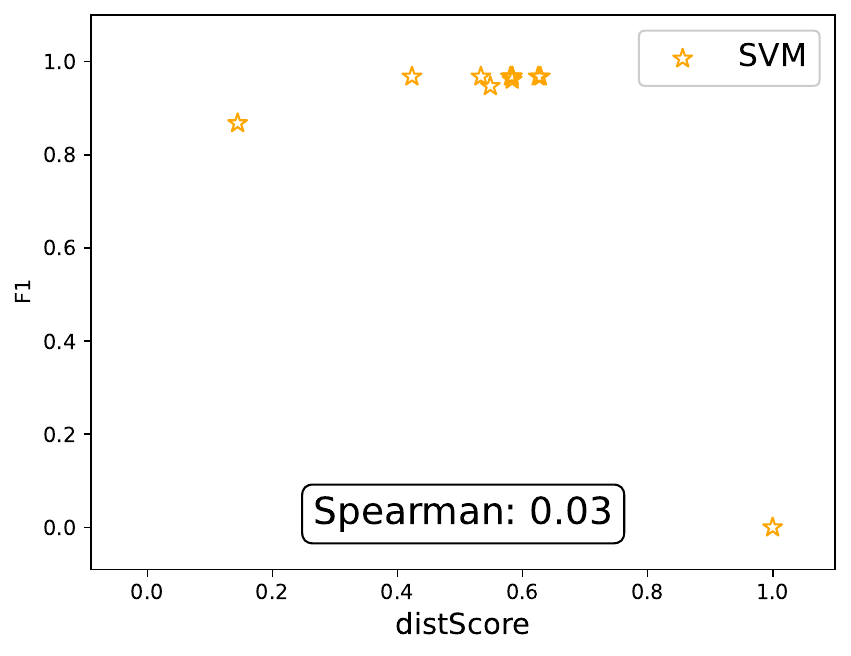}}\\

      & \raisebox{-0.5\height}{\includegraphics[width=0.29\textwidth] {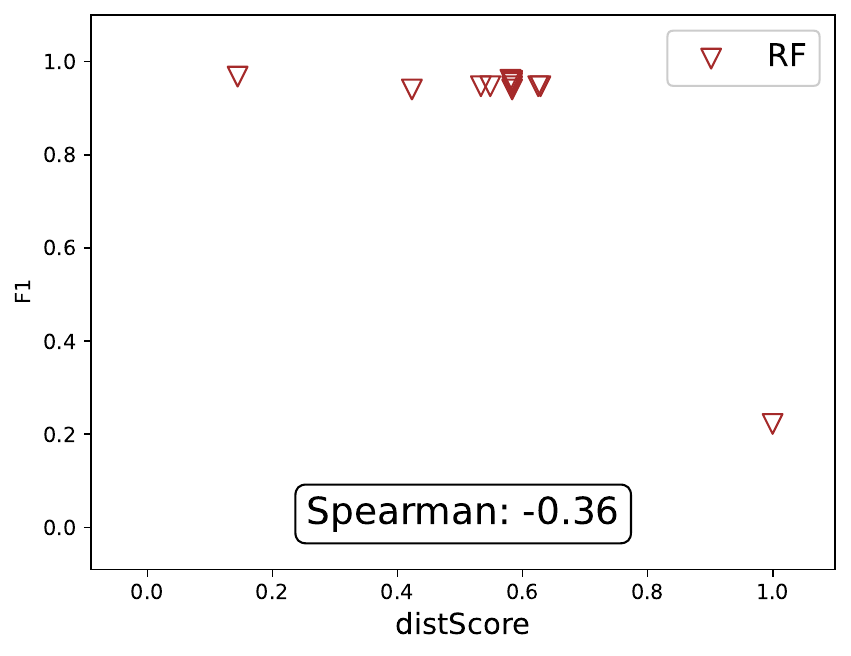}}

    \end{tabularx}

  \caption{\minrev{Relationship between $\mathit{distScore}$ and AD Accuracy (HDFS)}}
  \label{fig:rq2-hdfs-dist-ratio}
\end{figure}

\paragraph{HDFS dataset.}
\rev{\figurename~\ref{fig:rq2-hdfs-dist-ratio} shows the relationship between the degree of distinguishability (i.e., the $\mathit{distScore}$, shown in the x-axis) and the anomaly detection accuracy (i.e., the F1-score, shown in the y-axis) for the HDFS dataset.
Each sub-figure corresponds to a different anomaly detection technique, 
and each data point represents a log parsing technique.
\minrev{The Spearman correlation coefficient between the $\mathit{distScore}$ and the F1-score is also shown in each sub-figure.}
For DeepLog, LogAnomaly, LogRobust, and CNN, the F1-score mostly increases with the distinguishability score, except for an outlier around $\mathit{distScore} = 0.99$.
This means that the anomaly detection accuracy mostly improves when the log parsing results 
are more distinguishable, except for the outlier. 
This outlier is due to LogCluster, which generates 
an exceptionally high number of templates, \num{39998}, while the number of oracle templates is only 26 as noted in Table~\ref{table:datasets}.
\minrev{Although such a large number of templates leads to a high degree of distinguishability between 
normal and abnormal log sequences due to the high specificity of the templates,
it also leads to an excessive number of ``features'' to consider for the learning-based 
anomaly detection techniques, making the learning from training data more difficult, resulting in decreased anomaly detection accuracy.}
For the ML-based anomaly detection techniques, i.e., SVM and RF, the F1-score remains similar 
regardless of the distinguishability score, except for the same outlier discussed above.
We suspect that this is mainly because the traditional ML-based techniques are more sensitive 
to the number of features they use for learning (i.e., the number of templates, which typically range from 26 to 201) than to the degree of distinguishability.
However, LogCluster notably identifies a significantly higher number of templates, totaling \num{39998}.
For PLELog, the F1-score does not show a clear correlation with the distinguishability score.
This could be mainly due to the unique architecture of PLELog, 
which uses Gated Recurrent Units (GRUs) to model the log sequences, 
as discussed in Section~\ref{sec:eval-rq1-results}.

\begin{figure}
  \centering
  
\begin{tabularx}{\textwidth}{
          l X  @{\hspace{3pt}}  X@{\hspace{1pt}}  l 
    }

      & \raisebox{-0.5\height}{\includegraphics[width=0.29\textwidth] {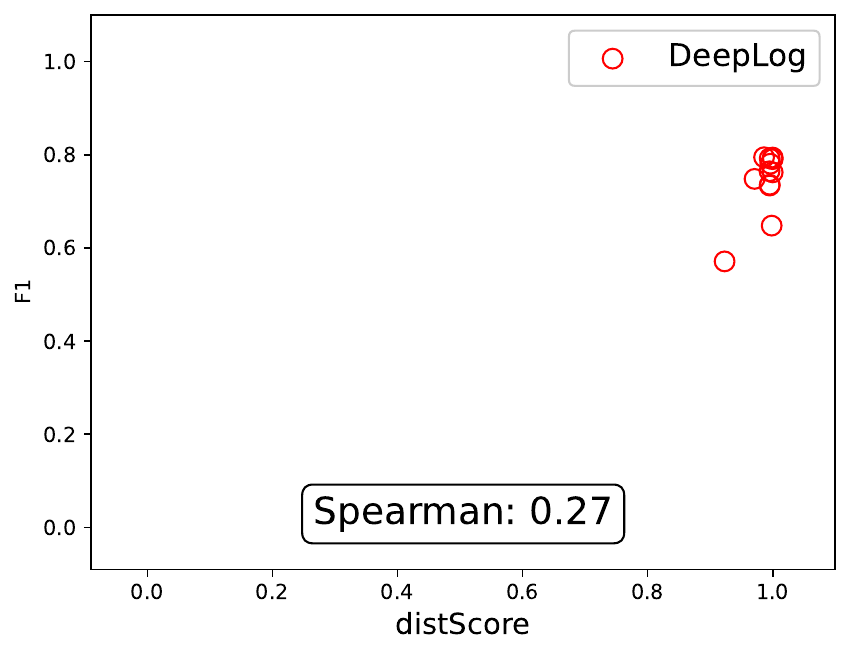}}
      & \raisebox{-0.5\height}{\includegraphics[width=0.29\textwidth] {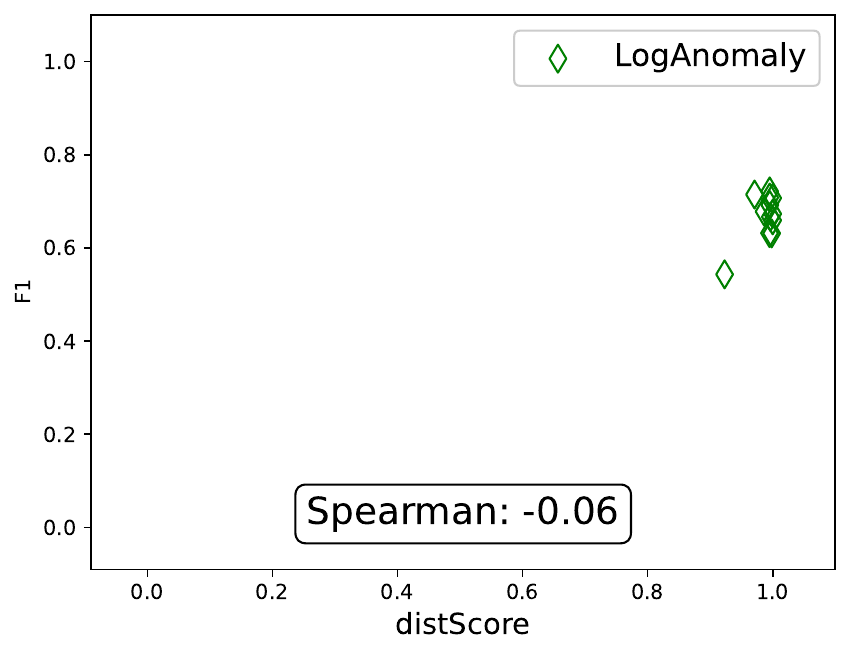}}
       & \raisebox{-0.5\height}{\includegraphics[width=0.29\textwidth] {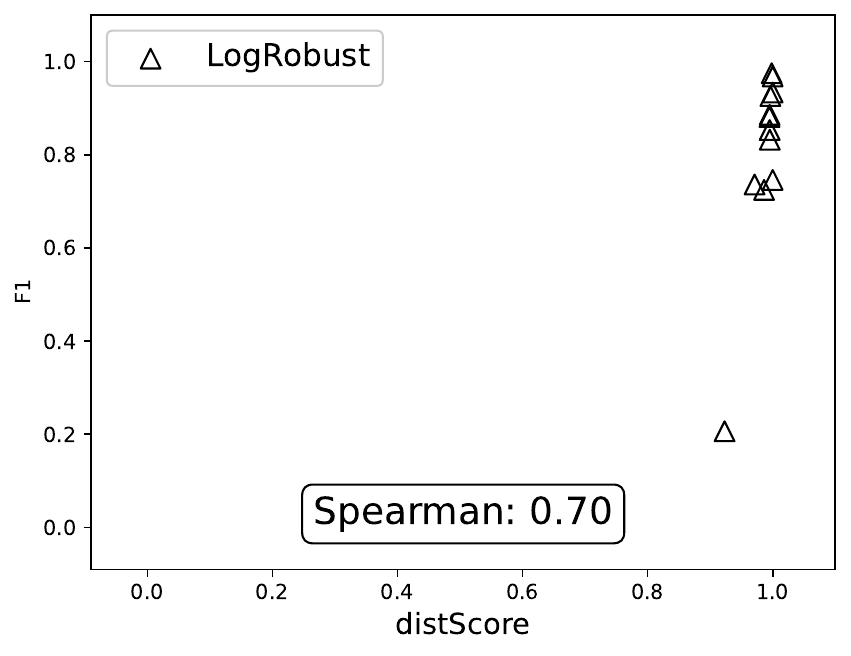}} \\
      
      &  \raisebox{-0.5\height}{\includegraphics[width=0.29\textwidth] {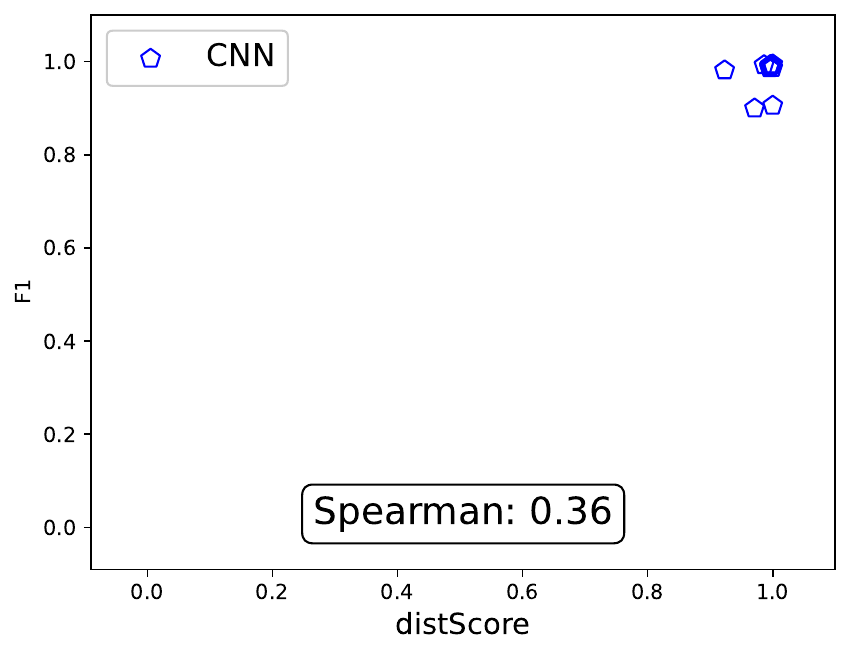}}
      &  \raisebox{-0.5\height}{\includegraphics[width=0.29\textwidth] {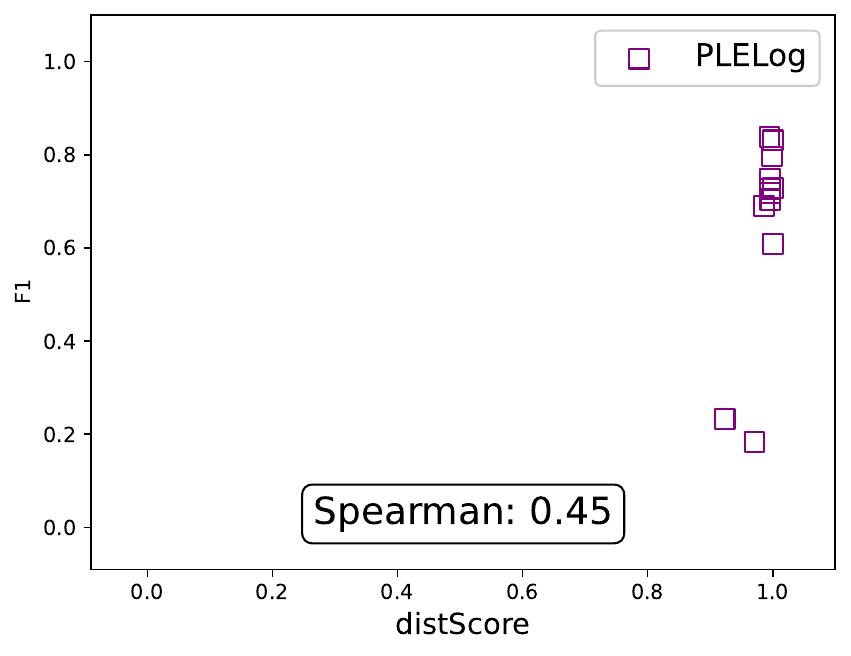}}
      &  \raisebox{-0.5\height}{\includegraphics[width=0.29\textwidth] {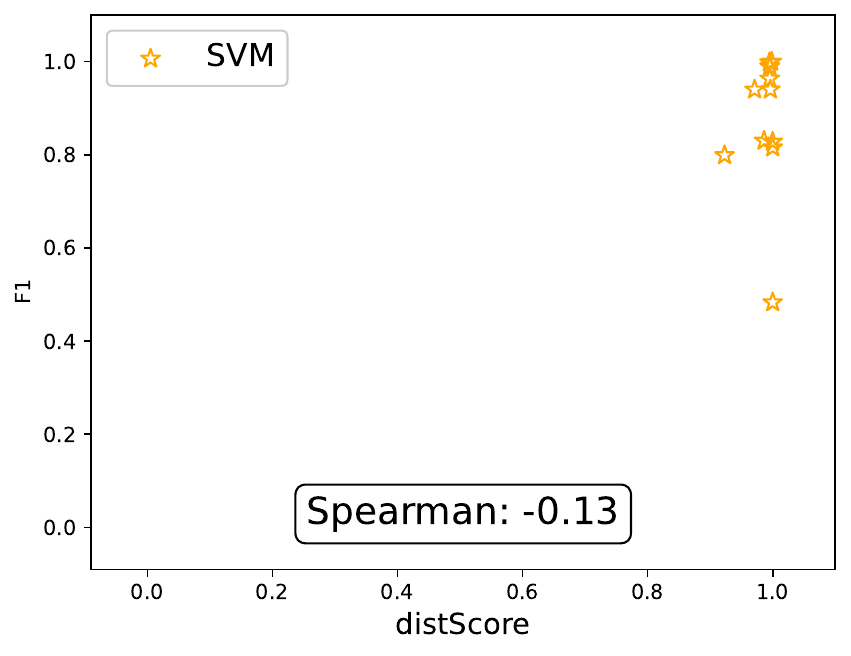}}\\

      & \raisebox{-0.5\height}{\includegraphics[width=0.29\textwidth] {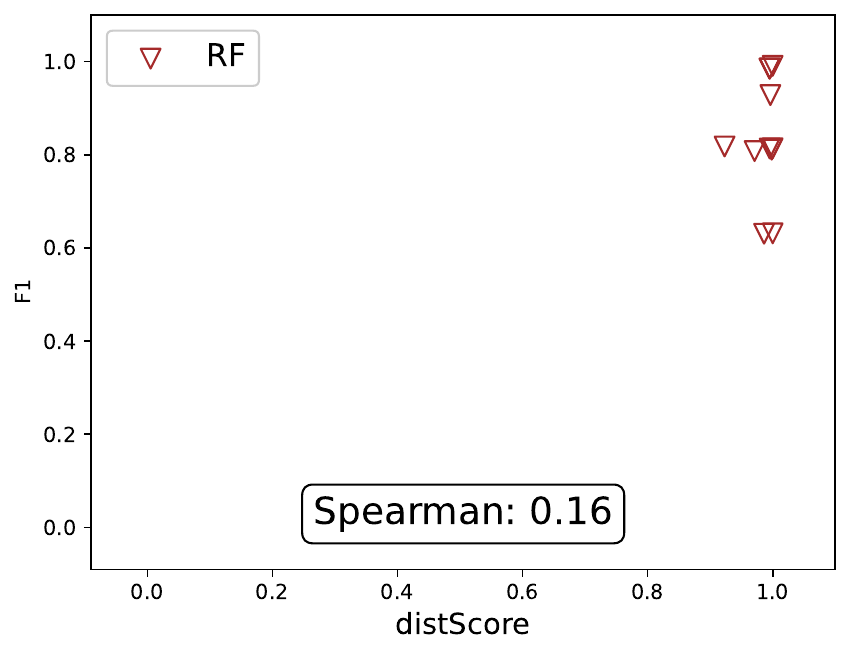}}

    \end{tabularx}

  \caption{\minrev{Relationship between $\mathit{distScore}$ and AD Accuracy (BGL)}}
  \label{fig:rq2-bgl-dist-ratio}
\end{figure}

\paragraph{BGL dataset.}
\minrev{\figurename~\ref{fig:rq2-bgl-dist-ratio} shows the results for the BGL dataset. 
The structure of the figure is the same as that of \figurename~\ref{fig:rq2-hdfs-dist-ratio}.
Overall, the F1-score mostly increases with the distinguishability score, except for LogAnomaly and SVM.
However, their Spearman correlations are very weak (only $-0.06$ and $-0.13$, respectively).
In other cases, the Spearman correlations are positive, ranging from 0.16 (RF) to 0.70 (LogRobust).
This implies that the findings from the HDFS dataset are generally consistent with those from the BGL dataset.}

To sum up, although the degree of distinguishability of log parsing results is not always 
positively related to anomaly detection accuracy, 
most of the deep learning-based techniques show moderate and positive correlations 
between the distinguishability degree and the anomaly detection accuracy.
Considering the heuristic nature of the proposed distinguishability score,
defining a more sophisticated and precise metric 
that can better capture the relationship between the distinguishability of log parsing results 
and the anomaly detection accuracy is an interesting direction for future work.}

\begin{tcolorbox}
\rev{The additional analysis for RQ2 shows that the degree of distinguishability of log parsing results is positively and moderately correlated with the accuracy of most deep learning-based anomaly detection techniques, but not for traditional machine learning-based techniques. 
This implies that distinguishability should be considered for deep learning-based log parsing for anomaly detection, and calls for defining more sophisticated metrics for measuring the degree of distinguishability.}
\end{tcolorbox}

\subsection{Threats to Validity}\label{sec:threats}

The used oracle templates determine log parsing accuracy 
values. For example, as noted by \citet{khan2022guidelines}, manually extracting 
oracle templates by investigating log messages without accessing the 
corresponding source code could result in biased, incorrect oracle templates. 
This could be a significant threat to the validity of our results. To mitigate 
this, we perused the source code (of the exact version that 
generated the logs) for each software system and used the templates directly 
extracted from the source code. Although this made us exclude a few log datasets 
whose source code was unavailable, it was beneficial to ensure the validity of 
our results. 

Individual log parsing and anomaly detection techniques have distinct 
hyper-parameters, which might significantly affect the log parsing and anomaly 
detection results. To mitigate this, we used the same hyper-parameter values 
proposed by the authors, when available; otherwise, we ran preliminary experiments  
and used the values that resulted in the same results reported in the corresponding papers. 

Using a specific set of log datasets is a potential threat to external
validity. Though the datasets we considered include the logs of various systems,
we had to select HDFS, Hadoop, and OpenStack due to the reasons discussed in Section~\ref{sec:datasets}.
Therefore, even though the datasets have been widely used
in existing literature~\cite{le2022log, chen2021experience} on
log-based anomaly detection, they may not capture diverse
characteristics of log data. Further experiments with different
datasets are required to improve the generalizability of our results.

In RQ2, we artificially generated pairs of distinguishable and indistinguishable log parsing results 
to systematically assess the impact of the distinguishability of log parsing results 
on anomaly detection accuracy using balanced data. 
To mitigate any bias introduced during the process, 
we carefully designed Algorithms~\ref{alg:indist} and \ref{alg:dist} 
to minimize the difference between each pair of log parsing results,
except for their distinguishability property.
Note that, although the pair generation process (by merging templates)
might look unrealistic, it reflects what frequently happens in real-world scenarios; 
for example, it is not uncommon for log parsing techniques to misidentify templates 
so that messages with different oracle templates are mapped to the same (misidentified) template.

 \section{Findings and Implications}\label{sec:implications}

One of the most surprising results from our evaluation is that, using 
all existing log parsing accuracy metrics in the literature, we did not find any significant correlation with anomaly detection accuracy. In other words, more 
accurate log parsing results are not necessarily better for anomaly detection 
accuracy. This implies that log parsing accuracy 
is not a good indicator of the quality of log 
parsing results for anomaly detection purposes. As explained with an example in 
Section~\ref{sec:eval-rq1-results}, this happens because inaccurate log parsing 
results can still be useful for anomaly detection as long as normal and abnormal 
logs are distinguishable. At the extreme, a log parsing result $R_{50}$ with 
50\% accuracy could be better for anomaly detection than a log parsing result 
$R_{100}$ with 100\% accuracy if $R_{50}$ distinguishes normal and abnormal logs 
while $R_{100}$ does not. 
\rev{This could happen when, for example, the log quality is poor
  (e.g., because of  inconsistencies between the developers' intentions and concerns on logging
  and the actual logging statements in the source code~\cite{9240659}) to
  the point that even using oracle templates cannot fully distinguish all normal log sequences from abnormal ones.}

This surprising finding leads to an important practical implication: When used 
for anomaly detection purposes, we can no longer choose a log parsing technique 
based on accuracy. Instead, as shown in Section~\ref{sec:eval-rq2-results}, 
the distinguishability of log parsing results should be the main selection criterion. For 
example, since normal and abnormal logs are often used for training anomaly detection 
models, candidate log parsing results should be compared in terms of their 
capability to distinguish normal and abnormal logs. If there are multiple 
techniques that can equally distinguish between normal and abnormal logs, then the one 
with the lowest number of identified templates would be preferred since reducing 
the number of templates would increase the performance of anomaly detection by 
reducing dimensionality (i.e., the number of features considered in machine 
learning models)~\cite{shin2021theoretical}. 

\minrev{Note that the notion of distinguishability for log parsing results is irrelevant if these results are not used for anomaly detection.
However, if anomaly detection needs log parsing (which is frequently the case in practice), then considering distinguishability can help engineers select the most suitable log parsing technique for anomaly detection.}

\rev{One may rightfully think that it is intuitive that the distinguishability of log parsing 
results is essential for learning-based anomaly detection techniques,
which distinguish between normal and 
abnormal log sequences by using the log parsing results (i.e., templates) as learning 
features. However, despite the prevalent use of log parsing in anomaly detection, the 
importance of distinguishability has been surprisingly ignored in the log analysis 
community. This paper aims to highlight the significance of distinguishability in log 
parsing for anomaly detection. Furthermore, this is the first work to empirically 
demonstrate the importance of distinguishability after the theoretical framework proposed by~\citet{shin2021theoretical}.}

Though our objective here is not to identify the ``best'' log parsing and anomaly 
detection techniques, through our experiments, we found that there 
is no single best technique that significantly outperforms the others in all 
cases. In the future, to develop better log parsing techniques targeting anomaly detection, it would 
beneficial to focus on distinguishability, which has not been the case so far.

 \section{Related Work}\label{sec:related-work}

\begin{table*}
\footnotesize
\caption{Comparison with related empirical studies}
\label{table:related_work}
\begin{tabularx}{\linewidth}{XXXX}
\toprule
      Category & \citet{le2022log} & \citet{fu2023empirical} & Our work   \\
 \midrule
 \rowcolor{gray!25}
Objective & Investigate different factors that might affect anomaly detection accuracy & 
Investigate the impact of log parsing techniques on anomaly detection accuracy & 
Evaluate the impact of log parsing accuracy and the distinguishability of log parsing results on anomaly detection accuracy \\ 
Log parsing accuracy metrics & N/A &PA &  PA, GA, and TA     \\
 \rowcolor{gray!25}
Oracle templates & N/A & Manually generated for 2K sample log messages & Extracted from the corresponding source code \\
Logs used for measuring log parsing accuracy & N/A & Only a small fraction of logs actually used for anomaly detection & All logs used for anomaly detection \\
 \rowcolor{gray!25}
Log parsing techniques & Drain, Spell, IPLoM and AEL & Drain, Spell, IPLoM, LFA, Logram, and LenMa  & Drain, Spell, IPLoM, AEL, LFA, Logram, LenMa, LogSig, LogCluster, LogMine, SHISO, MoLFI, and SLCT\\
Anomaly detection techniques & DeepLog, LogRobust, LogAnomaly, PLELog, and CNN & DeepLog, LogRobust, Principal Component Analysis (PCA), LogClustering, Logistic Regression (LR), and Decision Tree (DT) & DeepLog, LogRobust, LogAnomaly, PLELog, CNN, \rev{SVM, and RF}   \\
 \rowcolor{gray!25}
Distinguishability~\cite{shin2021theoretical} & Not considered & Not considered & Considered \\

\bottomrule
\end{tabularx}

 \end{table*}

Although individual techniques for log parsing and anomaly detection have been 
studied for a long time, systematic studies covering several techniques have 
only recently begun to emerge.
For example, the most comprehensive evaluation studies on many log parsing 
techniques~\cite{zhu2019tools,dai2020logram,khan2022guidelines} were conducted 
over the last four years. 
Similarly, the relationship between log parsing and anomaly detection has 
received little attention until very recently. Below, we summarize the recent 
studies related to this topic. 

\citet{shin2021theoretical} presented the first theoretical study considering 
the relationship between log parsing and anomaly detection. As described in 
Section~\ref{sec:distinguishability}, they established the concept of ideal log 
parsing results for anomaly detection. We adopted their theoretical foundation, 
especially the notion of \textit{distinguishability} in log parsing results, and 
empirically showed that distinguishability is indeed essential for anomaly detection. To the best of our knowledge, our work is the 
first empirical study showing the importance of log parsing distinguishability for anomaly detection. 

As explained in Section~\ref{sec:motivation}, \citet{le2022log} presented an empirical study on factors that could affect 
anomaly detection accuracy. 
Although a part of their study investigated the impact of log parsing on anomaly 
detection accuracy, they investigated four log parsing techniques but did not  
assess the impact of log parsing accuracy. As a result, they only showed that using 
different log parsing techniques leads to different anomaly detection
accuracy scores. In our 
study, on the other hand, we explicitly measured log parsing accuracy, collected 
160 pairs of log parsing accuracy and anomaly detection accuracy values using 
different combinations of log parsing and anomaly detection techniques, and showed that there is 
no strong correlation between log parsing accuracy and anomaly detection 
accuracy.

During the writing of this paper, \citet{fu2023empirical} also presented an 
empirical study on the impact of log parsing on anomaly detection performance. 
Although their motivation and research questions are close to ours, there are 
several key differences. 
First, for measuring log parsing accuracy, they used the manually generated, 
error-prone oracle templates~\cite{khan2022guidelines} provided with the 2K log 
messages randomly sampled by \citet{zhu2019tools}. In other words, only a very 
small fraction of the logs used for anomaly detection was used to measure log 
parsing accuracy in their study. In our study, however, the same logs used for 
anomaly detection are used to measure log parsing accuracy, and the oracle 
templates are directly extracted from the corresponding source code. 
Second, they considered only one log parsing accuracy metric (GA), whereas we 
considered all three log parsing metrics (GA, PA, and TA) since different metrics assess complementary aspects of log 
parsing~\cite{khan2022guidelines}. 
Third, log parsing distinguishability, which is an essential factor that substantially affects 
anomaly detection accuracy (as shown in our RQ2), is only considered
in our study.
Finally, they only considered two deep learning-based anomaly
detection techniques (DeepLog and LogRobust), and focused also on more
traditional machine learning approaches (such as Principal Component
Analysis, clustering, logistic regression, and decision trees).
Such differences allow us to report new findings and provide concrete recommendations, as 
summarized in Section~\ref{sec:implications}.

\citet{wu2023effectiveness} recently presented an empirical study
on the effectiveness of log representation for machine learning-based
anomaly detection. They considered different log representation
techniques, such as FastText~\cite{joulin2016fasttext}, Word2Vec~\cite{mikolov2013efficient}, 
TF-IDF~\cite{salton1988term} and
BERT~\cite{devlin2018bert}, used to convert textual log data into
numerical feature vectors for machine learning algorithms, such as
Support Vector Machine, Logistic Regression, Random Forest, 
CNN, and LSTM. As a part of their study, they investigated 
the impact of log parsing on anomaly detection 
when used with different log representation techniques 
(in particular, FastText and Word2Vec). 
The empirical results showed that, in general, 
using log parsing (i.e., Drain~\cite{he2017drain}) improves the quality of log
representations (over raw, unparsed data) and thereby the performance
of anomaly detection; they also reported that some models (e.g., CNN
and LSTM) are less sensitive to whether the log data is parsed or not, possibly due 
to the strong feature extraction and representation ability, and can offset the impact of 
noise generated by log parsing. 
In addition to these results, they also investigated 
the impact of additionally refining log parsing results using regular expressions 
and the impact of using different log parsing techniques.
The results showed that refining log parsing results do not significantly increase 
anomaly detection performance but using different log parsing techniques yields 
slight variations in anomaly detection performance.
However, for these additional investigations, they used only one anomaly 
detection technique (i.e., Logistic Regression) and two log parsing techniques 
(i.e., Drain~\cite{he2017drain} and LogPPT~\cite{le2023log}).
Furthermore, they did not study the relationship between 
log parsing accuracy and anomaly detection accuracy. On the contrary, 
we use 13 log parsing techniques and 5 DL-based anomaly detection techniques 
to comprehensively investigate the relationship between log parsing accuracy 
and anomaly detection accuracy. 

Table~\ref{table:related_work} summarizes the key differences between
the closely-related previous empirical studies (i.e., \citet{le2022log}, \citet{fu2023empirical}) and our work.

 \section{Conclusion and Future Work}\label{sec:conclusion-future-work}

In this paper, we reported on a comprehensive empirical study investigating the 
impact of log parsing on anomaly detection accuracy, using 13 log parsing 
techniques, five DL-based \rev{and two ML-based} anomaly detection techniques on \rev{three} publicly 
available log datasets. 
When analyzing log parsing results for anomaly detection, we were surprised not to find any significant relationship between log parsing accuracy and anomaly detection 
accuracy, regardless of  metric used for the former (including GA, PA, and FTA).
This implies that, as opposed to common research practice to date, we can no longer select a log parsing technique \rev{purely} based on its 
accuracy when used for anomaly detection.
Instead, we experimentally confirmed existing theoretical results showing that the distinguishability of log parsing results plays 
an essential role in achieving accurate anomaly detection. It is therefore highly 
recommended \rev{to consider distinguishability when utilizing log parsing results as input for anomaly detection.}

As part of future work, we plan to extend our study with more publicly available 
datasets and log parsing techniques~\cite{le2023log,tao2023parser},  
which were published during the writing of this paper,
to increase the generalizability of our results. 
We also aim to include state-of-the-art few-shot anomaly detection 
techniques~\cite{huang2022registration,pang2021explainable}, which require only 
a limited amount of training data and could be more effective in practice. We 
also plan to provide a more granular analysis of  distinguishability for log 
parsing results by defining a new metric that assesses the degree of 
distinguishability.
\rev{
Finally, we plan to assess the performance of anomaly detection
techniques that do not require log parsing~\cite{le2021log, mvula2023heart, nedelkoski2020selfAtt}.
}

\section*{Data Availability}\label{sec:data-availability}

The replication package of our empirical evaluation (including the Python implementations for log parsing techniques, anomaly detection techniques, helper scripts, and datasets) is available on Figshare~\cite{figshare}. 
\section*{Declarations}

\textbf{Funding and/or Conflicts of interests/Competing interests} The authors declare that they have no conflict of interest.
\newline\\
This  research  was  funded  in  whole, or  in  part,  by  the
  Luxembourg  National  Research Fund (FNR), grant reference C22/IS/17373407/LOGODOR. Lionel Briand was in part supported by the Canada Research Chair and Discovery Grant programs of the Natural Sciences and Engineering Research Council of Canada (NSERC), and the Science Foundation Ireland grant 13/RC/2094-2. For the purpose of open access, and in fulfillment of the
  obligations arising from the grant agreement, the authors have applied
  a Creative  Commons Attribution  4.0  International  (CC  BY  4.0) license  to  any  Author Accepted Manuscript version arising from this submission.

\bibliographystyle{spbasic.bst}

\end{document}